\documentclass[twocolumn,aps,prl,superscriptaddress]{revtex4-2}
\usepackage{amsmath,empheq}
\usepackage{amssymb}
\usepackage{mathrsfs}  
\usepackage{amsfonts}
\usepackage{booktabs}
\usepackage{physics}
\usepackage{graphicx} 
\usepackage{subfigure} 
\usepackage{epsfig}
\usepackage{color}
\usepackage{epstopdf}
\usepackage{xspace}
 \usepackage{rotating}
 \usepackage{appendix}
 \usepackage{chemformula}
 \usepackage{bm}
 \usepackage[colorlinks=true,linkcolor=blue,citecolor=blue,filecolor=blue,urlcolor=blue]{hyperref}

\newcommand{\ttJJ}{$t$-$t'$-$J$-$J'$\xspace}
\newcommand{\tJ}{$t$-$J$\xspace}

 \definecolor{bondiblue}{rgb}{0.0, 0.58, 0.71}

\begin{document}
\bibliographystyle{prsty}

 \title{Unconventional superconductivity near a flat band in organic and organometallic materials}
\author{Jaime Merino}
\affiliation{Departamento de F\'isica Te\'orica de la Materia Condensada, Condensed Matter Physics Center (IFIMAC) and
Instituto Nicol\'as Cabrera, Universidad Aut\'onoma de Madrid, Madrid 28049, Spain}
\author{Manuel Fern\'andez L\'opez}
\affiliation{Departamento de F\'isica Te\'orica de la Materia Condensada, Condensed Matter Physics Center (IFIMAC) and
Instituto Nicol\'as Cabrera, Universidad Aut\'onoma de Madrid, Madrid 28049, Spain}
\author{Ben J. Powell}
\affiliation{School of Mathematics and Physics, The University of Queensland, QLD 4072, Australia}

\begin{abstract}
We study electron correlation driven superconductivity on a decorated honeycomb lattice (DHL), which has a  low-energy flat band. On doping, we find singlet superconductivity with extended-$s$, extended-$d$ and $f$-wave symmetry mediated by magnetic exchange. $f$-wave singlet pairing is enabled by the  lattice decoration. The critical temperature is predicted to be significantly higher than on similar lattices lacking flat bands. We discuss how high-temperature superconductivity could be realized in the DHL materials such as \ch{Rb3TT. 2 H2O} and Mo$_3$S$_7$(dmit)$_3$. 
\end{abstract}
 \date{\today}

 \maketitle
 

The recent discovery of superconductivity in twisted bilayer graphene has lead to intense theoretical investigations of Cooper pairing 
in nearly flat band systems. The observation of superconductivity close to correlated
insulating states in twisted bilayer graphene \cite{jarillo1,jarillo2,balents2020} 
suggests that Coulomb repulsion  plays a
major role in its electronic properties including, possibly, superconductivity. 
Flat bands  enhance Coulomb scattering -- since scattering  
processes with any transferred momenta are allowed within the flat band -- 
leading to novel pairing states \cite{sayyad2020}.
Prior to the discovery of superconductivity in twisted bilayer graphene \cite{volovik2018}
flat band systems where proposed as a route towards room temperature 
superconductivity \cite{volovik2011a,volovik2011b} 
due to linear scaling of the critical temperature with the coupling. These considerations are very general and motivate the search for superconducting materials, beyond twisted bilayer graphene, with flat bands \cite{aoki2020}.

\begin{figure}
	\centering
	\includegraphics[width=\columnwidth,clip]{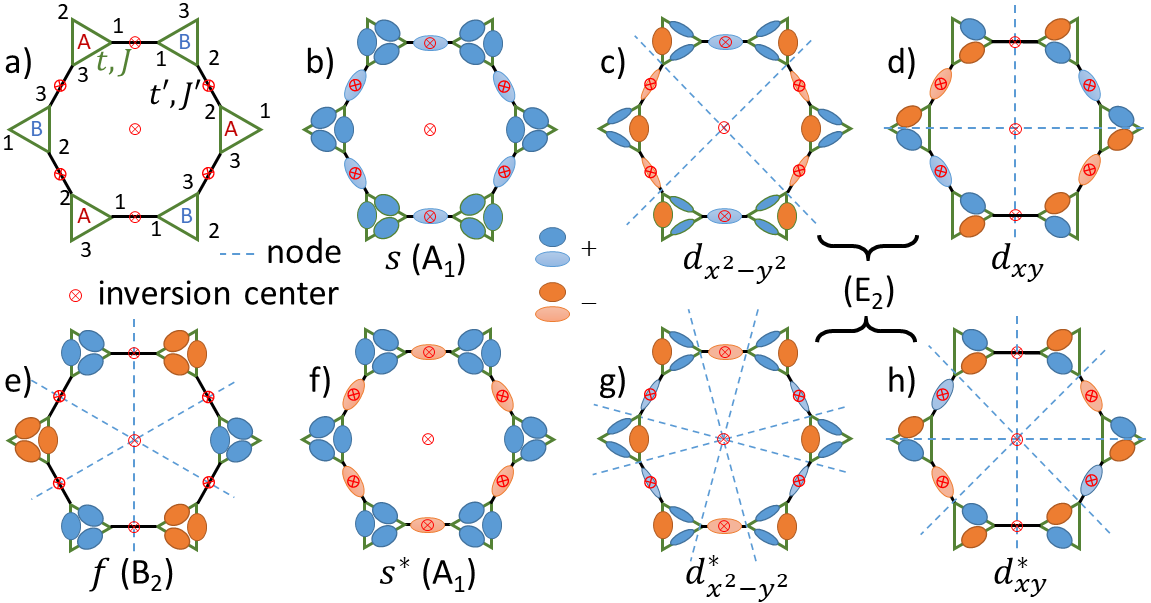}
	\caption{Decorated honeycomb lattice and its superconducting phases. (a) The decorated honeycomb lattice with the hopping ($t$, $t'$) and exchange ($J$, $J'$) parameters marked. The triangles form two sublattices (A and B) and contain three sites (labeled 1-3) (b-h) Real space representations of selected singlet superconducting states on the decorated honeycomb lattice. Color (size) of the ovals indicates the sign (magnitude) of the order parameter ($\Delta$, $\Delta'$). 
		The decorated lattice  allows an $f$-wave spin singlet state and extended-$s$ and -$d$ ($s^*$ and $d^*$) states 
	which are consistent with the 2D irreducible representation of $C_{6v}$ (Table \ref{tab:character}). The states in the bottom row (e-h) are found in our microscopic calculations.} 
	\label{fig:symmetry}
\end{figure}

\begin{figure}
	\centering
	\includegraphics[width=0.95\columnwidth,clip]{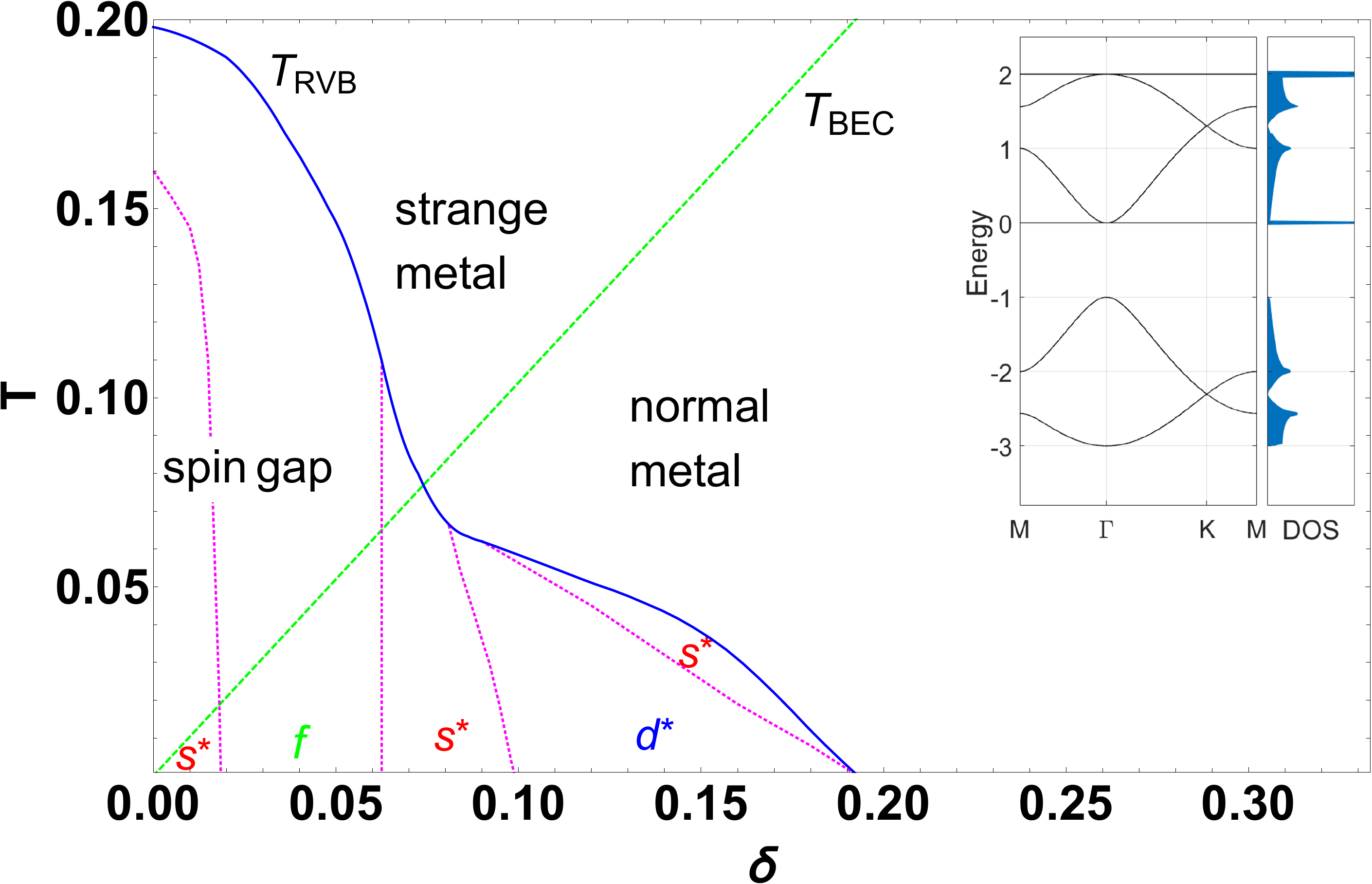}
	\caption{Phase diagram of the \ttJJ model on a decorated honeycomb lattice for $J/t=0.1$ and 
		$t'/t=J'/J=1$. 
		Pairing in  unconventional channels, $s^*$, $d^*$, and  
		$f$, occurs below the mean-field pairing temperature ($T_\text{RVB}$; blue solid line). Charge transport is coherent  below the Bose-Einstein condensation 
		temperature, $T_\text{BEC}$. Superconductivity requires both pairing and coherence, i.e., $T<T_\text{RVB}$ and $T_\text{BEC}$. The dotted (magenta) lines are first order transitions between different
		superconducting states while the solid (blue) line corresponds to a second order transition 
		between superconducting and metallic states. 
		Inset: The non-interacting band structure 
		of the decorated honeycomb lattice displaying flat bands and the corresponding DOS for $t'=t$.
	$t=1$ in all plots. 
	}
	\label{fig:phased}
\end{figure}

Electrons on the decorated honeycomb lattice (DHL; Fig. \ref{fig:symmetry}a) can display many interesting properties including topological phases  \cite{ruegg,wen,scarola2018,manuel2019,manuel2020}.
The  DHL has a flat band at the Fermi energy when half-filled and lightly hole doped [$1\leq\delta<1/3$, where the number of electrons per site is $n=1-\delta$;  Fig. \ref{fig:phased} (inset)]. This leads to large density of
states so one expects strong electronic correlation
effects  close to half-filling \cite{nourse2020}.
The DHL is realized in several materials including trinuclear organometallic compounds
\cite{khosla,jacko,merino,powell,Llusar}, organic molecular crystals \cite{Agawa},
iron(III) acetates \cite{iron3}, coordination polymers/metal-organic frameworks (MOFs)
\cite{henline,henling}, and  cold fermionic atoms  in
optical lattices \cite{cdmft}. An important open question is whether superconductivity from Coulomb interaction can arise in these DHL compounds as theoretically predicted in bare honeycomb systems \cite{chubukov2014,scalettar2018}.


Numerical work \cite{honecker2004,orus2018,manuel2020} indicates that the ground state of the spin-$1/2$ Heisenberg model on a DHL 
with nearest-neighbor interactions  is a valence bond solid (VBS).
Two different VBSs are suggested depending on the anisotropy ($J'/J$; Fig. \ref{fig:symmetry}a).
Subtle changes in the lattice and interactions can radically change this -- indicating that there are many competing ground states. For example,  the ground state of the Kitaev model on the DHL
is also a  quantum spin liquid (QSL)  \cite{kivelson2007}.  
A QSL is also predicted for the 
Heisenberg model on the kagom\'e lattice \cite{elser1989,lhuilier1997,singh2007,white2010}, which is closely related to the DHL. 
In contrast, longer range interactions and/or 
higher order spin exchange are needed to stabilize a QSL on the (anisotropic) triangular lattice \cite{motrunich2005,holt2014,merino2014,kenny2020}.  

Anderson's resonating valence bond (RVB) hypothesis \cite{anderson1987} is that unconventional superconductivity can arise when one dopes holes into a Mott insulator as valence bonds become mobile (singlet) Cooper pairs. 
Therefore, if VBSs occur at half-filling on the DHL \cite{honecker2004,orus2018,manuel2020}, then an important issue is to 
understand the conducting phases that arise upon hole doping. Similar programs have been carried out  in the context of the (square lattice) cuprate superconductors \cite{anderson1987,lee2006} and the (anisotropic triangular lattice) organic superconductors \cite{powell05,powell07}. Due to the greater complexity of the DHL 
we can expect, in general,  pairing states other than  
the $d$ and $d+id$ states generally found on the square and anisotropic triangular lattices \cite{powell05,powell07,lee2006,scalapino2012}. 

Here, we report on the existence of unconventional superconductivity, including 
an $f$-wave singlet state (Fig. \ref{fig:symmetry}e), in the hole doped DHL arising from
Coulomb repulsion. Successive  transitions from extended-$s$ ($s^*$) to  extended-$d$ ($d^*$) to $f$-wave superconductivity
occur at  low temperatures (Fig. \ref{fig:phased}).
The highest critical temperature
in our phase diagram occurs around $(7-8)\%$ doping where $s^*$ superconductivity is most
favorable. The superconducting critical  and  pseudogap temperatures are 
much larger than the corresponding ones on the square \cite{kotliar1988} and triangular lattices  \cite{lee2004,shastry2003,powell05,powell07}. Hence, the robustness 
of the superconducting and pseudogap phases is correlated with the 
flat band at the Fermi energy in the DHL.  

At first glance, an $f$-wave singlet state seems to violate the requirement that the wavefunction must be antisymmetric under the exchange of two fermions (electrons). Usually one thinks that if the wavefunction is odd under spatial inversion ($\bm k\rightarrow-\bm k$) then it must be even under  spin inversion ($\sigma\rightarrow-\sigma$); thus, as $f$-wave state must be a spin triplet. But, this argument does not account for the internal degrees of freedom within the unit cell of the DHL -- which can be described as either the site labels or as molecular orbital degrees of freedom \cite{khosla}. 

Insight into $f$-wave singlet states can be gained from  writing a (non-superconducting) two-electron wavefunction, $|\Psi_-\rangle$, that is odd under both spatial and spin inversion for a single unit cell of the DHL.  Let $|\Psi_\alpha\rangle = ( h_{\alpha 1, \alpha 2}^\dagger + h_{\alpha 2, \alpha 3}^\dagger + h_{\alpha 3, \alpha 1}^\dagger )|0\rangle$, where the singlet operator $h_{\alpha i, \beta j}^\dagger={1 \over \sqrt{2} } (c^\dagger_{\alpha i\uparrow} c^\dagger_{\beta j\downarrow}-c^\dagger_{\alpha i\downarrow} c^\dagger_{\beta j\uparrow} )$ and $c^{\dagger}_{\alpha i \sigma}$ creates an electron with spin $\sigma$, on the $i$th site of the $\alpha$th triangle. Define $|\Psi_\pm\rangle \equiv |\Psi_A\rangle \pm |\Psi_B\rangle$: both wavefunctions are a superposition of singlets within the triangles (and therefore singlets themselves) and satisfy fermionic antisymmetry for any pair of electrons, but whereas $|\Psi_+\rangle$ is even under inversion, $|\Psi_-\rangle$ is odd \cite{suppl}. The $f$-wave singlet superconducting state in Fig. \ref{fig:symmetry} is highly analogous to $|\Psi_-\rangle$.

 

Our microscopic theory considers the \ttJJ model on the DHL:
$
H = -t \sum_{\langle \alpha i, \alpha j \rangle \sigma} P_G ( c^\dagger_{\alpha i \sigma} c_{\alpha j \sigma}  + c^\dagger_{\alpha j \sigma} c_{\alpha i \sigma} ) P_G
- t' \sum_{\substack{\langle A i, B i \rangle, \sigma }} P_G ( c^\dagger_{A i \sigma} c_{B i\sigma}  + c^\dagger_{B i \sigma} c_{A i \sigma} ) P_G  
- J \sum_{\langle \alpha i, \alpha j \rangle } ( {\bf S}_{\alpha i} \cdot {\bf S}_{\alpha j} - {1 \over 4} n_{\alpha i} n_{\alpha j} ) 
- J'\sum_{\substack{\langle A i, B i \rangle }} ( {\bf S}_{A i} \cdot {\bf S}_{B i} - {1 \over 4} n_{A i} n_{B i} )
- \mu \sum_{\alpha i \sigma} c^\dagger_{\alpha i \sigma} c_{\alpha i \sigma},
$
where ${\bf S}_{\alpha i}=\sum_{\sigma\sigma'}c^{\dagger}_{\alpha i \sigma} {\bm\tau}_{\sigma\sigma} c_{\alpha i \sigma'}$, ${\bm\tau}$ is the vector of Pauli matrices, $n_{\alpha i}=\sum_{\sigma}c^{\dagger}_{\alpha i \sigma}  c_{\alpha i \sigma}$, and the Gutzwiller projector $P_G=\Pi_i (1- n_{i\uparrow} n_{i\downarrow} )$ excludes 
doubly occupied sites. 
The sums are restricted to nearest-neighbor sites either within a triangle, $\langle \alpha i, \alpha j\rangle $ or between neighboring triangles, $\langle A i, B i \rangle $, cf. Fig. \ref{fig:symmetry}a. 
Motivated by superexchange, and to reduce the number of free parameters, we set $J'/J=(t'/t)^2$ in all of our calculations. We solve this model via RVB theory \cite{lee2006}, where double occupancy is projected out of a Bardeen-Cooper-Schrieffer (BCS) wavefunction via the Gutzwiller approximation (GWA). Technical details are given in \cite{suppl}.

The DHL has $C_{6v}$ symmetry, 
which has six irreducible representations (Table \ref{tab:character}), three even and three odd under inversion symmetry [in 2D, inversion is equivalent to a C$_2$ rotation about the `$z$-axis':  both map $(x,y)\rightarrow(-x,-y)$].  
The order parameters most relevant to our RVB calculations are sketched in Fig. \ref{fig:symmetry}. This includes a $B_2$ ($f_{x(3y^2-x^2)}$, henceforth $f$-wave)  state built from a superposition of singlets. This state is odd under inversion through the center of the dodecahedron ($\cong C_2$) and under inversion through the center of the $t'$-bonds (equivalent to a $\sigma_v$ mirror). But, it is even under the $\sigma_d$ mirrors (which bisect the triangles). This is consistent with fermionic exchange statistics because the singlets are all within a single triangle that decorate the honeycomb lattice -- the nodes lie on the inter-triangle bonds, along the $\sigma_v$ mirror planes. $p$-wave  singlets are also possible on this lattice via a similar construction (with nodes along the $t'$ bonds) \cite{footnote}. However, the $p$-wave  states are not low-energy solutions in our RVB theory.


\begin{table}
	\centering	
	\begin{tabular}{l|cccccc|c}   
					& E 	& 2C$_6$	& 2$C_3$		& $C_2$	& 3$\sigma_v$	& 3$\sigma_d$	& superconducting order	\\ \hline
		$A_{1}$		& 1 	& 1			& 1				& 1		& 1				& 1				& $s$, $s^*$	\\
		$A_{2}$		& 1 	& 1			& 1				& 1		& -1			& -1			& $i_{xy(3x^4-10x^2y^2+3y^4)}$	\\
		$B_{1}$		& 1 	& -1		& 1				& -1	& 1				& -1			& $f_{y(3x^2-y^2)}$	\\
		$B_{2}$ 	& 1 	& -1		& 1				& -1	& -1			& 1				& $f_{x(3y^2-x^2)}$	\\
		$E_{1}$		& 2 	& 1			& -1			& -2	& 0				& 0				& $(p_x,p_y)$	\\
		$E_{2}$		& 2 	& -1		& -1			& 2		& 0				& 0				& $(d_{x^2-y^2},d_{xy})$, $(d_{x^2-y^2}^*,d_{xy}^*)$	\\
	\end{tabular}
	\caption{
		Character table for $C_{6v}$. The $n$-fold rotations (C$_n$) are about the center of the dodecahedron, $\sigma_d$ reflections are defined to pass through the center of $t'$-bonds, and $\sigma_v$ are  reflections through the line joining a vertex and the center of a triangle, cf. Fig. \ref{fig:symmetry}. The last column gives the conventional name of the superconducting symmetry.
	}\label{tab:character}
\end{table}

Odd-parity $f$- and $p$-wave singlet states are not allowed on the honeycomb lattice because the triangles are replaced by a single site. And singlets cannot form within a single site due to the strong Coulomb repulsion. Thus, the decorated lattice structure is directly responsible for allowing odd-parity singlet superconductors. 
Similarly, the $s^*$ state sketched in Fig. \ref{fig:symmetry}f does not have a natural analogue on the honeycomb lattice as the triangles are replaced by a single site. Clearly, similar superconducting states should be possible on other decorated lattices. We stress that the $f$-wave singlet is not an artifact of the 2D model and that this construction works equally well for 3D decorated lattices.



When hole doped, the model displays unconventional superconductivity, 
 Fig. \ref{fig:phased}. Below the mean-field temperature, $T_\text{RVB}$, unconventional Cooper pairing is stabilized by the spin exchange interactions. 
 %
Superconductivity  occurs when the
 preformed Cooper pairs  Bose condense, {i.e.}
  when $T<T_\text{RVB}$ and $T<T_\text{BEC}$. In quasi-two-dimensional systems \cite{shastry2003} 
 $T_\text{BEC} \approx {1 \over 2 + \ln( 4 \gamma \over \pi) } { \delta \over \rho^* } \simeq 1.04 \delta$ for the condensation of  bosons 
 at the bottom of the lowest DHL band (where $\rho^* =0.14$ is
 the density of states and $\gamma=100$, quantifies the 
 large anisotropy of the dispersion perpendicular to 
 the lattice plane  \cite{shastry2003}).
 We find that $T_c=T_\text{BEC} = T_\text{RVB}\approx0.075 t$ at $\delta_c \approx 0.075$ (optimal doping) for $t'/t=J'/J=1$. The  superconducting critical temperature $T_c=T_\text{BEC}$ for $\delta \leq \delta_c$;  whereas for $T_c=T_\text{RVB}$ for $\delta \geq \delta_c$.
 In comparison, for the optimally doped \tJ 
 model on the triangular lattice $T_c \sim 0.017 t$ \cite{lee2004}, 
 suggesting that the flat band in the DHL significantly enhances $T_c$.

Below $T \approx 0.16 t$ we find an RVB state with $s^*$ symmetry for the undoped DHL; this is consistent with exact diagonalization \cite{manuel2020,suppl} on small clusters. 
Thus we find that $s^*$ superconductivity emerges at small hole 
doping in the DHL from a parent insulating state with $s^*$ character.
This is highly analogous to the way $d_{x^2-y^2}$ superconductivity arises from hole doping a parent state with $d_{x^2-y^2}$ character in the \tJ model on the square 
lattice \cite{lee2006}.


On further doping the system at $T \rightarrow 0$ the pairing symmetry changes at a series of first order superconductor-superconductor transition.
For $0.023 \lesssim \delta \lesssim 0.06$ $f$-wave pairing occurs;
$d^*$ pairing is stabilized in the range $0.1 < \delta < 0.2$; and for other $\delta$, $s^*$ pairing is again present.
Above $\delta>0.22$ the metallic state is recovered. Our numerical analysis finds degenerate $d_{xy}^*$ and $d_{x^2-y^2}^*$ solutions  (with the same \textit{free} energy). This contrasts with the 
$d_{x^2-y^2} +i d_{xy}$ solution found in honeycomb \cite{doniach2007} and triangular lattices \cite{baskaran2003,shastry2003,ogata2003}. We note that these two different types of $d^{(*)}$
solutions (degenerate and $d^{(*)}+id^{(*)}$) are 
both expected from 
the symmetry of the (decorated) honeycomb lattice: the 
Ginzburg-Landau theory of the 
$d^{(*)}$-order parameters belonging to the E$_2$ 2D irreducible representation, Table I, predicts a  $d_{x^2-y^2}^{(*)} +i d_{xy}^{(*)}$ state  in the weak coupling limit, but can also give degenerate $d_{xy}^{(*)}/d_{x^2-y^2}^{(*)}$ states away from the BCS limit \cite{ben2006}.

The metallic phases also have unconventional properties \cite{anderson1987,baskaran1987,kotliar1988,shastry2003}.
For instance, a pseudogap phase  with a spin gap (but no charge gap) emerges for $\delta < \delta_c$ and $T_\text{BEC}<T<T_\text{RVB}$. In this phase one expects a dip in the density of states at the Fermi level, in contrast to the peak observed in conventional Fermi liquids. In the present flat band system, our analysis shows that the pseudogap phase 
is stable  to much higher temperatures than in other lattices without flat bands. 
For $J=0.1 t$ and $\delta \rightarrow 0$, we find that the pseudogap temperature $T^*=T_\text{RVB} \sim 2J$. In contrast  $T^* \sim 0.75 J$ on the square lattice \cite{kotliar1988} and $T^* \sim 0.2 J$  on the triangular lattice \cite{lee2004,shastry2003}.
A strange metallic phase is expected to occur 
in the intermediate doping range for $T>T_\text{RVB}$ and $T_\text{BEC}$, with a Fermi liquid predicted for $T_\text{BEC}>T>T_\text{RVB}$ \cite{lee2006}.

\begin{figure}[t!]
	\centering
	\includegraphics[width=0.9\columnwidth,clip]{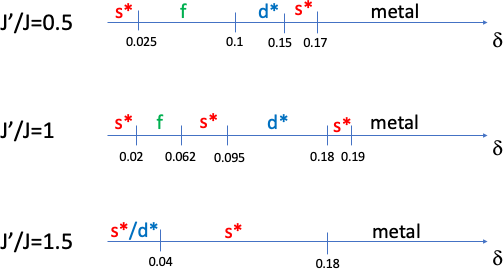}
	\caption{Dependence of pairing states on exchange anisotropy. 
		For $J'<J$ the $f$-wave phase is enlarged (relative to $J'=J$) and the $d^*$-wave phase is reduced. 
		For $J'>J$ the $s^*$ phase dominates with the $d^*$-wave state only energetically competitive for small
doping and the $f$-wave phase absent. The approximate critical dopings, $\delta$, for the
various transitions are displayed. We have fixed $J/t=0.1$, $T/t=0.01$.
	}
	\label{fig:phase-diag-delta}
\end{figure}

Anisotropic interactions lead to dramatic changes in the symmetry of the Cooper pairing. 
Low temperature phase diagrams 
as a function of doping for different $J'/J=(t'/t)^2$ 
are shown in Fig. \ref{fig:phase-diag-delta}. 
Small $J'/J$ increases the range of dopings in which the $f$-wave singlet phase is stable, but suppresses $d^*$ pairing dramatically. 
In contrast, at large $J'/J$, the $f$-wave solution is no longer realized; and $s^*$ pairing dominates the phase diagram. However, at very low doping the $s^*$ and $d^*$ solutions become quasi-degenerate, i.e., so close in free energy that we cannot reliably determine which is the lowest energy solution.
Thus, while $f$-wave singlet pairing is more likely to occur at low hole doping and $J'/J \lesssim 1$, $d^*$ pairing would typically arise at larger hole doping.

\begin{figure}[t!]
	\centering
	\includegraphics[width=4.25cm,clip]{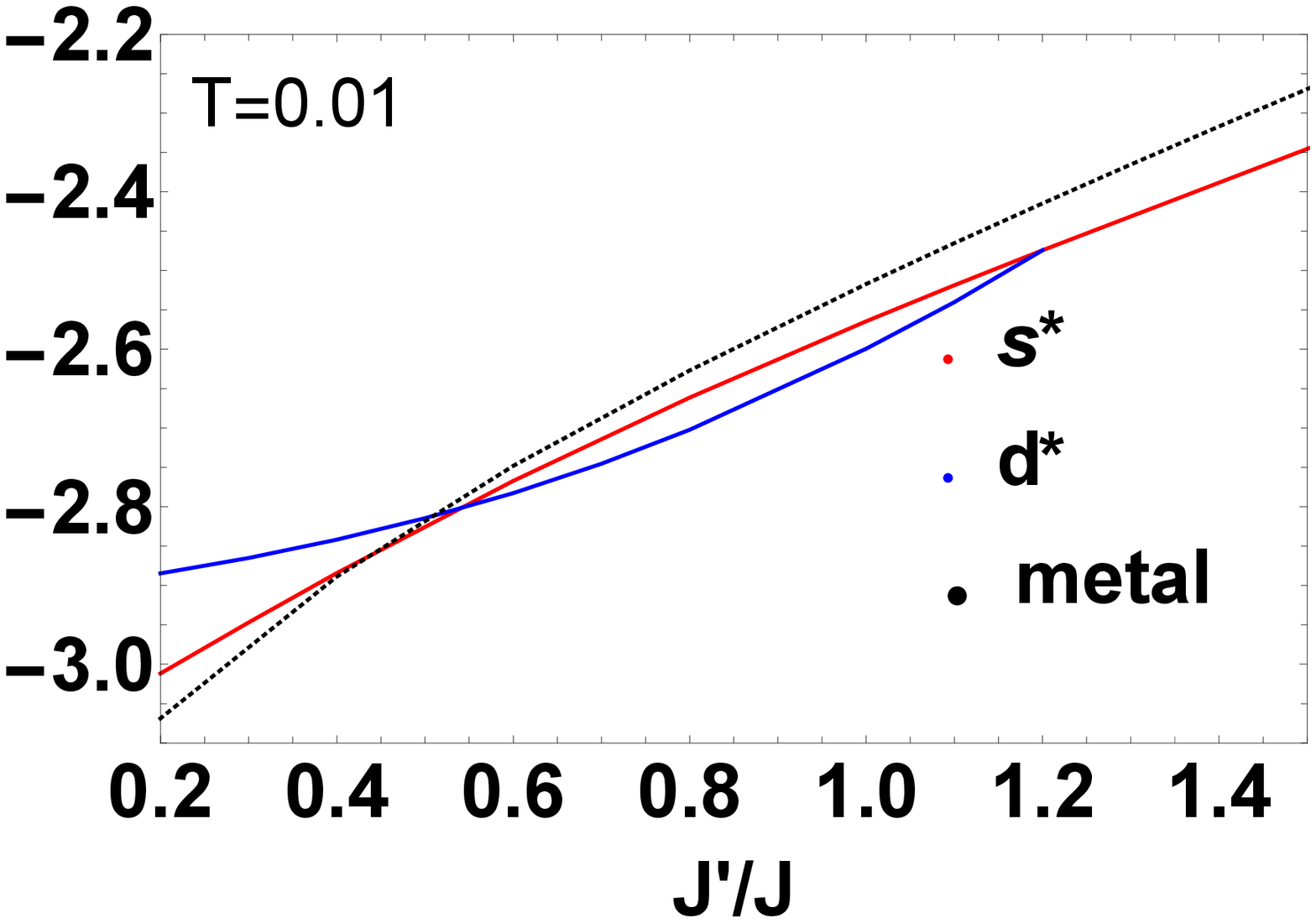}
	\includegraphics[width=4.25cm,clip]{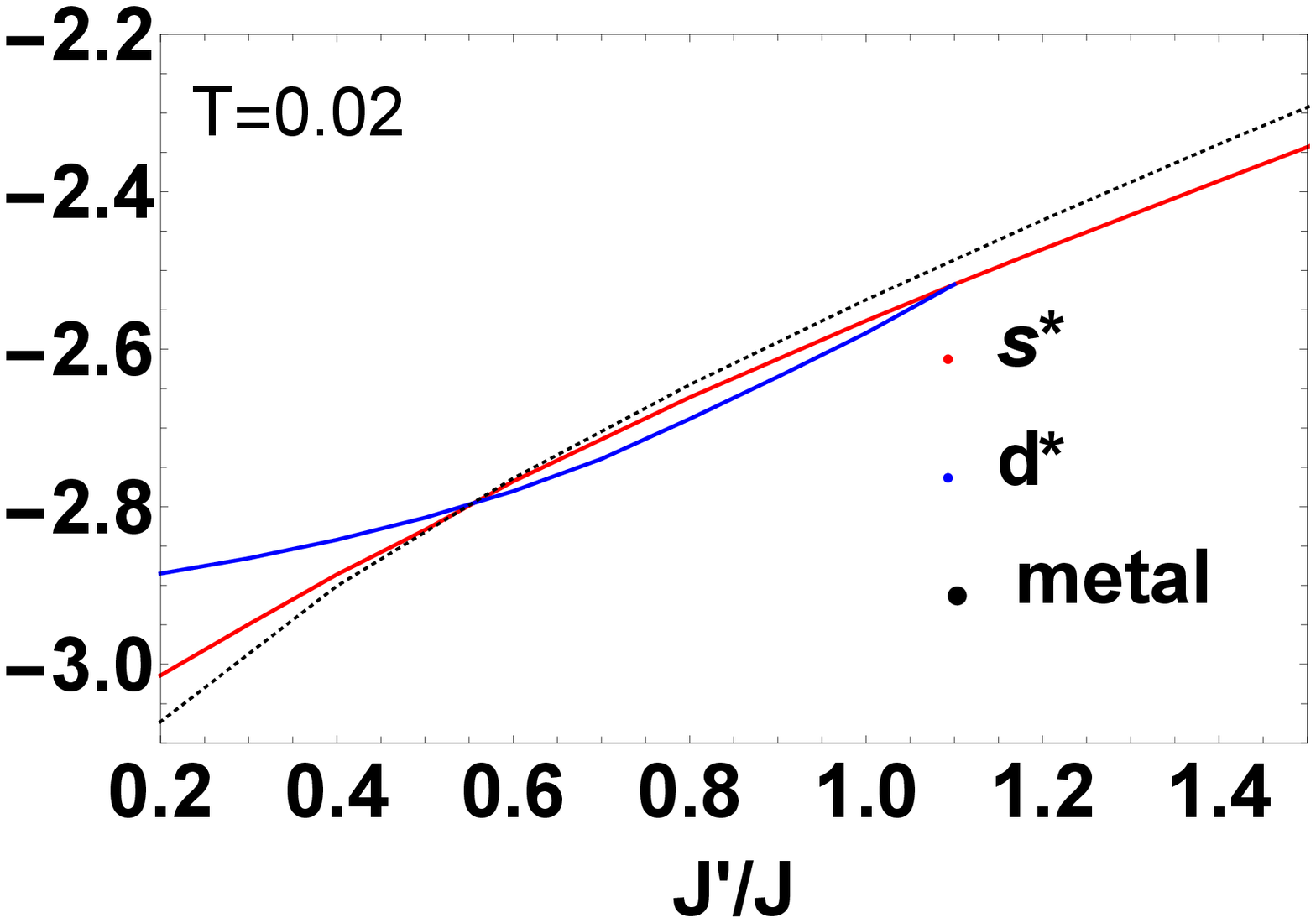}
	\includegraphics[width=4.25cm,clip]{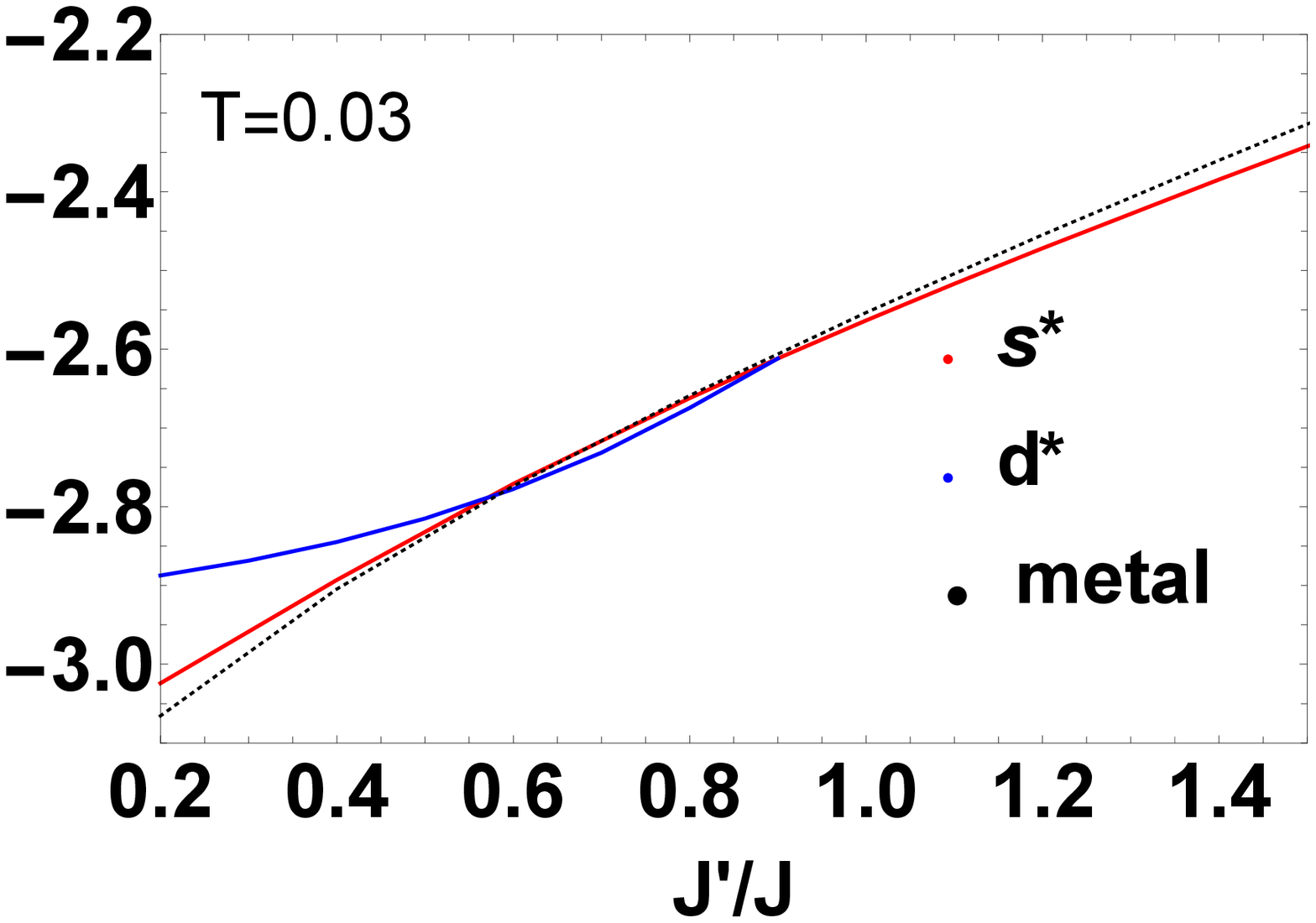}
	\includegraphics[width=4.25cm,clip]{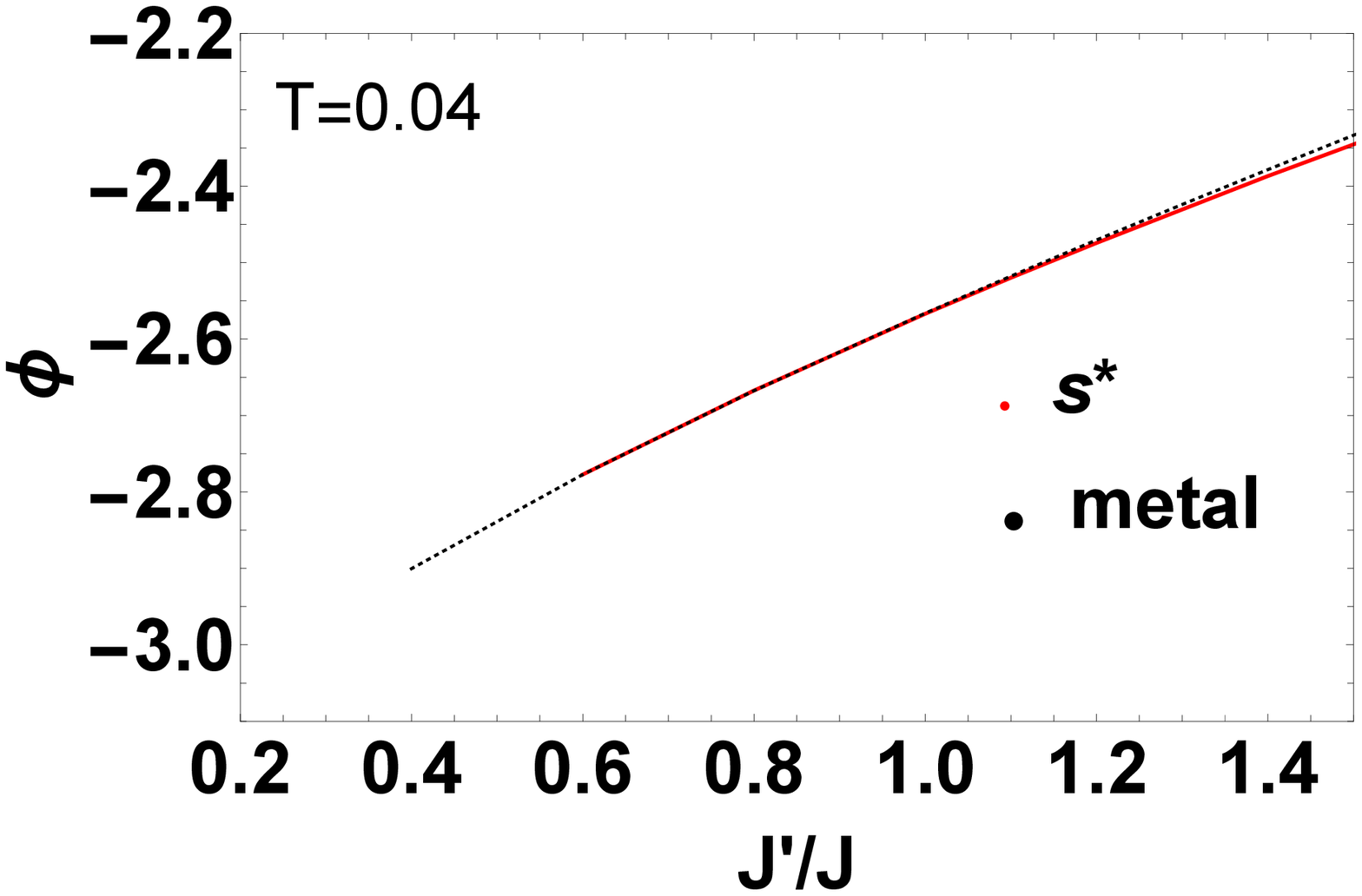}
	\caption{Dependence of superconducting and metallic free energies on $J'/J=(t'/t)^2$ at fixed doping, $\delta=0.15$ and $J/t=0.1$. The free energies at different temperatures of the $s^*$ and $d^*$ pairing states are compared to the metallic solution. Depending on $J'/J$ either $s^*$ or $d^*$ pairing occurs at low  temperatures. For $T \gtrsim 0.04t$, superconductivity disappears giving way to a metal for almost any $J'/J$.  }
	\label{fig:E_v_aniso}
\end{figure}

Raising the temperature also favors $s^*$ pairing. 
For example we compare the free energies of the lowest energy superconducting solutions to the metallic state for $\delta=0.15$ in  Fig. \ref{fig:E_v_aniso}. 
At low temperatures, $d^*$ pairing occurs over a broad range of exchange anisotropy, $0.5< J'/J <1.2$. 
The $d^*$-wave solution is much more rapidly suppressed by thermal fluctuations than $s^*$ pairing. Thus at these larger $\delta$ we 
expect an $s^*$ superconductor immediately below $T_c$, followed by a  transition to a $d^*$ state at lower $T$, cf. Fig. \ref{fig:phased}. 

Significant insight into the temperature and doping variations of the pairing symmetry can be gained from  analyzing a phenomenological weak coupling  \ttJJ model where  doubly occupied sites are not
projected out  \cite{suppl}. 
The linearized gap equations predict nine 
possible superconducting states: $s^{(*)},f, p_x, p_y, d^{(*)}_{x^2-y^2}$, and $d^{(*)}_{xy}$ all 
contained in Table. \ref{tab:character}, as expected. 
In general, the different solutions have  different 
$T_c$'s. The complicated dependence of the $T_c$'s on the coupling, $g$, indicates that transitions between different superconducting states  occur on increasing $g$ (Fig. S5 \cite{suppl}). The 
GWA projection  effectively amounts to renormalizing the parameters of the \ttJJ model: $(J/t, J'/t) \rightarrow ({\tilde J}/{\tilde t}, {\tilde J'}/{\tilde t})={2 \over \delta (1+\delta) } (J/t, J'/t)$. Since the effective coupling, $g={\tilde J}/{\tilde t}$, increases as $\delta \rightarrow 0$, different superconducting states can, in principle, be stabilized. 

The weak coupling  \ttJJ model allows us to make contact 
with previous work on superconductivity in 
graphene by taking the limit of $t'/t, (J'/J) \rightarrow 0$. In this limit, the DHL band structure equivalent to two copies of the homeycomb lattice plus two flat bands with a large seperation between the molecular orbital of the triangles \cite{jacko2015}. 
Previous work on an unprojected \tJ model on the honeycomb lattice finds $d$-wave superconductivity \cite{doniach2007}.
In contrast, in the above limit of our fully projected \ttJJ model, we find quasi-degenerate $s^*$ and $f$-wave superconductivity with a transition to a metallic state occurring at a rather small hole doping, $\delta_c \sim 0.045$ for $J'/J=0.1$  ($t'/t \sim 0.326$). 
Apart from the fact that there are no analogues of the $s^*$ and $f$-wave states 
considered here (Fig. \ref{fig:symmetry}) on the honeycomb lattice, our analysis on the unprojected \ttJJ model for, say, $J'/J=0.1$,  shows 
that $d^*$ pairing is the most favorable solution at weak coupling (Fig. S5 \cite{suppl}) in agreement with \cite{doniach2007}. Since the effective couplings,
${\tilde J}/{\tilde t},{\tilde J'}/{\tilde t'}$ in our projected \ttJJ model increase as $\delta \rightarrow 0$
we would have expected that the system goes from 
$d^*$ pairing at large doping (small effective couplings)
to $s^*/f$ pairing at small doping.
However, our numerical calculations in the projected \ttJJ model 
show that, as doping is increased, the metallic state sets in before the $d^*$ solution 
is stabilized. This explains why we do not observe 
$d^*$ superconductivity in our 
projected \ttJJ model on the DHL when $J'/J \rightarrow 0$.

Thus, we have seen that flat bands in the DHL provide a route to exotic, high-temperature superconductivity. 
In particular we predict unconventional $s^*$, $d^*$, and $f$-wave singlet superconducting states competing with one another due to the complex structure of the lattice which produces the flat band. 
The flat band at the Fermi energy enhances both the superconducting critical temperature and the pseudogap temperature scale compared to other lattices studied with comparable theories.
To quantify this, let us assume that the doped DHL can be experimentally realized in organic or organometallic materials. For Mo$_3$S$_7$(dmit)$_3$ (dmit=1,3-dithiol-2-thione-4,5-dithiolate) $t' \approx t=0.05$~eV \cite{jacko2015}, taking $J'=J=0.1 t$  yields $T_c \sim 44$~K at optimal doping. 
In the half-filled  insulating 
material \ch{Rb3TT. 2 H2O} (TT=triptycene tribenzoquinone) the largest hopping integral is an order of magnitude greater than the largest hopping integral in Mo$_3$S$_7$(dmit)$_3$ \cite{Agawa}, suggesting, surprisingly, that superconductivity may survive to even higher temperatures of the order of $\sim 10^2$ K.
In \ch{Rb3TT. 2 H2O} the  $t$ and $t'$ are negative
so the flat band lies below the Fermi energy and electron (rather than hole) doping promises flat band superconductivity; this might be achieved via the synthesis of Rb$_{3-\delta}$Sr$_\delta$\ch{TT. 2 H2O}. 




\begin{acknowledgments}
	We thank Henry Nourse and Ross McKenzie for helpful conversations. We acknowledge financial support from (Grant No. RTI2018-098452-B-I00) MINECO/FEDER, Uni\'on Europea, from the Mar\'ia de Maeztu Programme for Units of Excellence in R\&D 
	(Grant No. CEX2018-000805-M), and the Australian Research Council (Grant No. DP180101483).
\end{acknowledgments}



\begin{thebibliography}{30}
\bibitem{footnote} An $f_{y(3x^2-y^2)}$ state constructed from singlets along the $t'$ bonds (with nodes along the $t$ bonds) is also possible, but would treble the size of the unit cell.
\end{thebibliography}
\end{document}


\bibliographystyle{prsty}

 \title{Supplementary material for unconventional superconductivity near a flat band in organometallic materials}
  \author{Jaime Merino}
\affiliation{Departamento de F\'isica Te\'orica de la Materia Condensada, Condensed Matter Physics Center (IFIMAC) and
Instituto Nicol\'as Cabrera, Universidad Aut\'onoma de Madrid, Madrid 28049, Spain}
\author{Manuel Fern\'andez L\'opez}
\affiliation{Departamento de F\'isica Te\'orica de la Materia Condensada, Condensed Matter Physics Center (IFIMAC) and
Instituto Nicol\'as Cabrera, Universidad Aut\'onoma de Madrid, Madrid 28049, Spain}
\author{Ben J. Powell}
\affiliation{School of Mathematics and Physics, The University of Queensland, QLD 4072, Australia}
 \date{\today}
\maketitle

\section{Hartree-Fock-Bogoliubov decoupling}

Using bond singlet operators \cite{baskaran2002}: $h_{\alpha i, \beta j}^\dagger={1 \over \sqrt{2} } (c^\dagger_{\alpha i\uparrow} c^\dagger_{\beta j\downarrow}
-c^\dagger_{\alpha i\downarrow} c^\dagger_{\beta j\uparrow} )$, the exchange terms in model (1) of the
main text can be expressed as: $\sum_{\langle \alpha i, \beta j \rangle} ({\bf S}_{\alpha i} \cdot {\bf S}_{\beta j} - {n_{\alpha i} n_{\alpha j} \over 4})= -\sum_{\langle \alpha i, \beta j \rangle} h_{\alpha i ,\beta j}^\dagger h_{\alpha i, \beta j}$, where sums are restricted to nearest-neighbor sites in the lattice.

\begin{figure}
   \centering
\includegraphics[width=4cm]{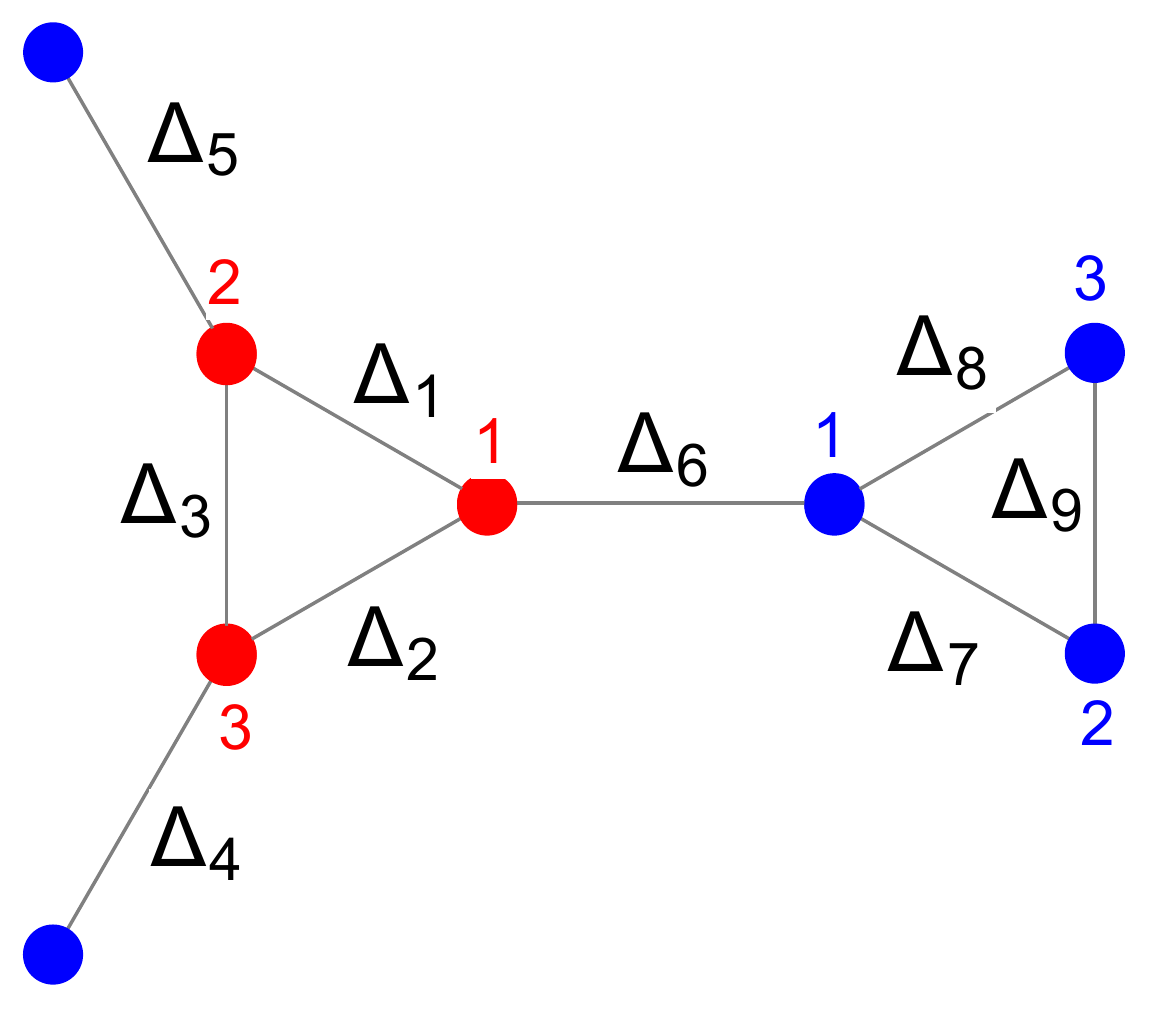}
\caption{Real space Cooper pairing amplitudes. All the pairing amplitudes encoded in the $9$-dimensional 
vector, ${\bf \Delta }$, used in the mean-field analysis and linearized gap equations. The two triangular clusters in the unit cell are shown in red (A) and blue (B).}
 \label{fig:pairs}
 \end{figure}

We perform a Hartree-Fock-Bogoliubov decoupling of the exchange terms introducing mean-field order parameters:
\begin{eqnarray}
\chi_{\alpha i, \beta j}  &=& \langle c^\dagger_{\alpha i, \sigma} c_{\beta j, \sigma} \rangle,
\nonumber  \\
\Delta_{\alpha i, \beta j}  &=& \langle h_{\alpha i,\beta j} \rangle.
\end{eqnarray}
Hence, the mean-field hamiltonian, $H_{MF}=H_t+H_J$ in momentum space reads:
\begin{widetext}
\begin{eqnarray}
H_t &=& -t\sum_{\alpha{\bf k}\langle i, j \rangle,\sigma} \left( c^\dagger_{\alpha i\sigma}({\bf k}) c_{\alpha j\sigma} ({\bf k}) + h. c.  \right)-t' \sum_{\substack{{\bf k}, \langle i, j \rangle, \sigma} }\left( e^{i {\bf k}\cdot 
{\bf \delta}_{ij} } c^\dagger_{A i\sigma}({\bf k}) c_{B j\sigma} ({\bf k}) + h.c. \right)  
\nonumber \\
H_J &=& -J\sum_{\alpha{\bf k}\langle i, j \rangle,\sigma} \left( {\chi_{\alpha i, \alpha j} \over 2} c^\dagger_{\alpha i\sigma}({\bf k}) c_{\alpha j\sigma} ({\bf k}) + h. c.  \right)- J'\sum_{\substack{{\bf k}, \sigma, \langle i, j \rangle} }\left( {\chi_{A i, B j} \over 2}  e^{i {\bf k}\cdot \delta_{ij}} c^\dagger_{A i\sigma} ({\bf k}) c_{B j\sigma}({\bf k})  + h.c. \right)  
\nonumber \\
&-&J \sum_{\alpha{\bf k}\langle i,j \rangle}  \Delta_{\alpha i,\alpha j} \left( c^\dagger_{\alpha i\uparrow}({\bf k}) c^\dagger_{\alpha j\downarrow} (-{\bf k}) - c^\dagger_{\alpha  i\downarrow}({\bf k}) c^\dagger_{\alpha j\uparrow} (-{\bf k})  \right ) +h.c.
\nonumber \\
&-& J'\sum_{{\bf k}\langle i,j \rangle} \Delta_{A i,B j} \left( c^\dagger_{A i\uparrow}({\bf k}) c^\dagger_{B j\downarrow} (-{\bf k}) -c^\dagger_{A i\downarrow}({\bf k}) c^\dagger_{B j\uparrow} (-{\bf k})  \right) + h.c.
\nonumber \\
&+& J \sum_{\langle \alpha i, \alpha j \rangle } \left( |\Delta_{\alpha i,\alpha j}|^2 +|\chi_{\alpha i, \alpha j}|^2 \right) +J' \sum_{\langle A i, B j \rangle} \left( |\Delta_{A i,B j}|^2 + |\chi_{A i, B j}|^2  \right)+\mu \sum_{\alpha} c^\dagger_{\alpha\sigma} ( {\bf k}) c_{\alpha\sigma} ({\bf k}),
\end{eqnarray}
\end{widetext}
with $i,j=1,2,3$ enumerating the sites inside each triangular unit and $\alpha, \beta=A, B$ referring to any two nearest-neighbor triangular units of the lattice. Thus the sums are restricted
to n.n. sites in the same triangle, $\langle \alpha i , \alpha j \rangle$
 or n.n. sites between two neighbor triangles, $\langle A i, B j \rangle $. The two neighbor $A, B$ triangles with the different pairing amplitudes entering the HF-BdG decomposition of the model are shown in Fig. \ref{fig:pairs}. 

The diagonalized hamiltonian reads:
\begin{eqnarray}
H_{MF} &=& \sum_{m,{\bf k}, \sigma} \omega_m({\bf k} ) \left( \gamma^\dagger_{m \sigma} ({\bf k}) \gamma_{m \sigma} ({\bf k}) - {1 \over 2}  \right) 
\nonumber \\
&+& J\sum_{\langle \alpha i, \alpha j \rangle } (|\Delta_{\alpha i,\alpha j}|^2+|\chi_{\alpha i, \alpha j}|^2) 
\nonumber \\
&+& J'\sum_{\substack{\langle A i, B j \rangle} } (|\Delta_{A i,B j}|^2+|\chi_{A i,B j}|^2),
\nonumber \\
\end{eqnarray}
where $\omega_m({\bf k})>0$ with $m=1,...,6$ are the positive Bogoliubov quasiparticle dispersions. 

The free energy of the system reads:
\begin{eqnarray}
\Phi &=& -{1 \over \beta} \sum_{m,{\bf k}, \sigma} \ln(1+ e^{-\beta \omega_m({\bf k})} )- \sum_{m,{\bf k}} \omega_m({\bf k}) + 6 N_s \mu 
\nonumber \\
&+& J \sum_{\langle \alpha i, \alpha j \rangle } \left( |\Delta_{\alpha i,\alpha j}|^2 +|\chi_{\alpha i, \alpha j}|^2 \right)
\nonumber \\
&+& J' \sum_{\substack{\langle A i, B j \rangle} } \left( |\Delta_{A i,B j}|^2 + |\chi_{A i, B j}|^2  \right).
\end{eqnarray}
From the minimization of the free energy we arrive at the set of self-consistent equations (SCE):
\begin{eqnarray}
\chi_{\alpha i,\alpha j} &=& -{1 \over J N_s} \sum_{m,{\bf k},\sigma} \left( f(\omega_m ({\bf k}) ) -{1 \over 2} \right) \left( {\partial \omega_m({\bf k}) \over \partial \chi^*_{\alpha i, \alpha j} }\right),
\nonumber \\
\chi_{A i, B j} &=& -{1 \over J' N_s} \sum_{m,{\bf k},\sigma} \left( f(\omega_m ({\bf k}) ) -{1 \over 2} \right) \left( {\partial \omega_m({\bf k}) \over \partial \chi^*_{A i, B j} }\right),
\nonumber \\
\Delta_{\alpha i,\alpha j} &=& -{1 \over J N_s} \sum_{m,{\bf k},\sigma} \left( f(\omega_m ({\bf k}) ) -{1 \over 2} \right) \left( {\partial \omega_m({\bf k}) \over \partial \Delta^*_{\alpha i, \alpha j} }\right),
\nonumber \\
\Delta_{A i,B j} &=& -{1 \over J' N_s} \sum_{m,{\bf k},\sigma} \left( f(\omega_m ({\bf k}) ) -{1 \over 2} \right) \left( {\partial \omega_m({\bf k}) \over \partial \Delta^*_{A i, B j} }\right),
\nonumber \\
\delta &=& -{1 \over 6 N_s} \sum_{m,{\bf k}, \sigma} \left( f(\omega_m({\bf k}))- {1 \over 2} \right) \left( {\partial \omega_m({\bf k})\over \partial \mu} \right),\notag\\
\label{eq:sce}
\end{eqnarray}
which are solved numerically. The doping is defined as $\delta=1-n$, where $n$ is the average electron occupancy per site. 

Note that the final GWA projected solution of the $t-t'-J-J'$ model effectively implies substituting $(J, J') \rightarrow ({\tilde J}, {\tilde J'})=(g_J J, g_{J'} J')$ and $(t, t') \rightarrow ( {\tilde t}, {\tilde t'})=(g_t t, g_{t'} t')$ with $g_t=g_{t'}= {2 \delta \over 1 + \delta}$ and $g_{J}=g_{J'}= {4 \over (1 + \delta )^2}$,
in the free energy and the above set of self-consistent equations.

\section{Superconducting critical temperatures and pairing symmetries}

Following \cite{pickett2018} the critical temperatures can be obtained from the set of linearized gap equation:
\begin{equation}
\Gamma {\bf \Delta} = {1 \over J} {\bf \Delta},
\label{eq:eigen}
\end{equation}
where ${\bf \Delta}$ is the nine component vector describing the allowed real space pairing amplitudes
and is defined as follows: 
\begin{widetext}
\begin{equation}
{\bf \Delta}=(\Delta_{A1,A2}, \Delta_{A1, A3}, \Delta_{A2, A3},
\Delta_{A3,B3}, \Delta_{A2,B2}, \Delta_{A1,B1}, \Delta_{B1,B2}, \Delta_{B1,B3}, \Delta_{B2,B3})^T,
\label{eq:pairs}
\end{equation}
with the real space pairing amplitudes taken as in Fig. \ref{fig:pairs}.
The $9 \times 9$ matrix $\Gamma$ reads:
\begin{equation}
\Gamma_{pq}={1 \over 2 N_s} \sum_{m,n,{\bf k}} \tanh({\epsilon_m ({\bf k}) \over 2 k_B T} ) {1 \over \epsilon_m({\bf k}) +\epsilon_n({\bf k})} \left( \Omega^{*p}_{m n} ({\bf k})
\Omega^{q}_{mn}({\bf k})+\Omega^{p}_{mn}({\bf k})\Omega^{*q}_{mn}({\bf k}) \right),
\end{equation}
\end{widetext}
obtained from the SCE equations (\ref{eq:sce}) by expanding the quasiparticle energies: 
\begin{equation}
 \omega_m({\bf k})=\epsilon_m({\bf k})+\sum_{n} { |\Delta_{m,n}({\bf k})|^2 \over \epsilon_m({\bf k}) + \epsilon_n({\bf k})},
 \end{equation}
where, $\epsilon_m({\bf k})$ are the eigenvalues of the kinetic energy part of the hamiltonian, $H_t$. The pairing amplitudes read
$\Delta_{m,n}({\bf k})=\sum_{p} \Omega^p_{mn}({\bf k}) J_p \Delta_p$,
where the $J_p$
encode the exchange couplings corresponding to the real space pairing amplitudes, $\Delta_p$, introduced in Eq. (\ref{eq:pairs}) and shown in Fig. \ref{fig:pairs}, i.e.,
$J_p=J$ for $p=1,2,3$ and $p=7,8,9$ (spin exchange inside the two triangular clusters) while 
$J_p=J'$ for $p=4,5,6$ (spin exchange between the two clusters). 
\begin{figure*}[!t]
   \centering
\includegraphics[width=8cm]{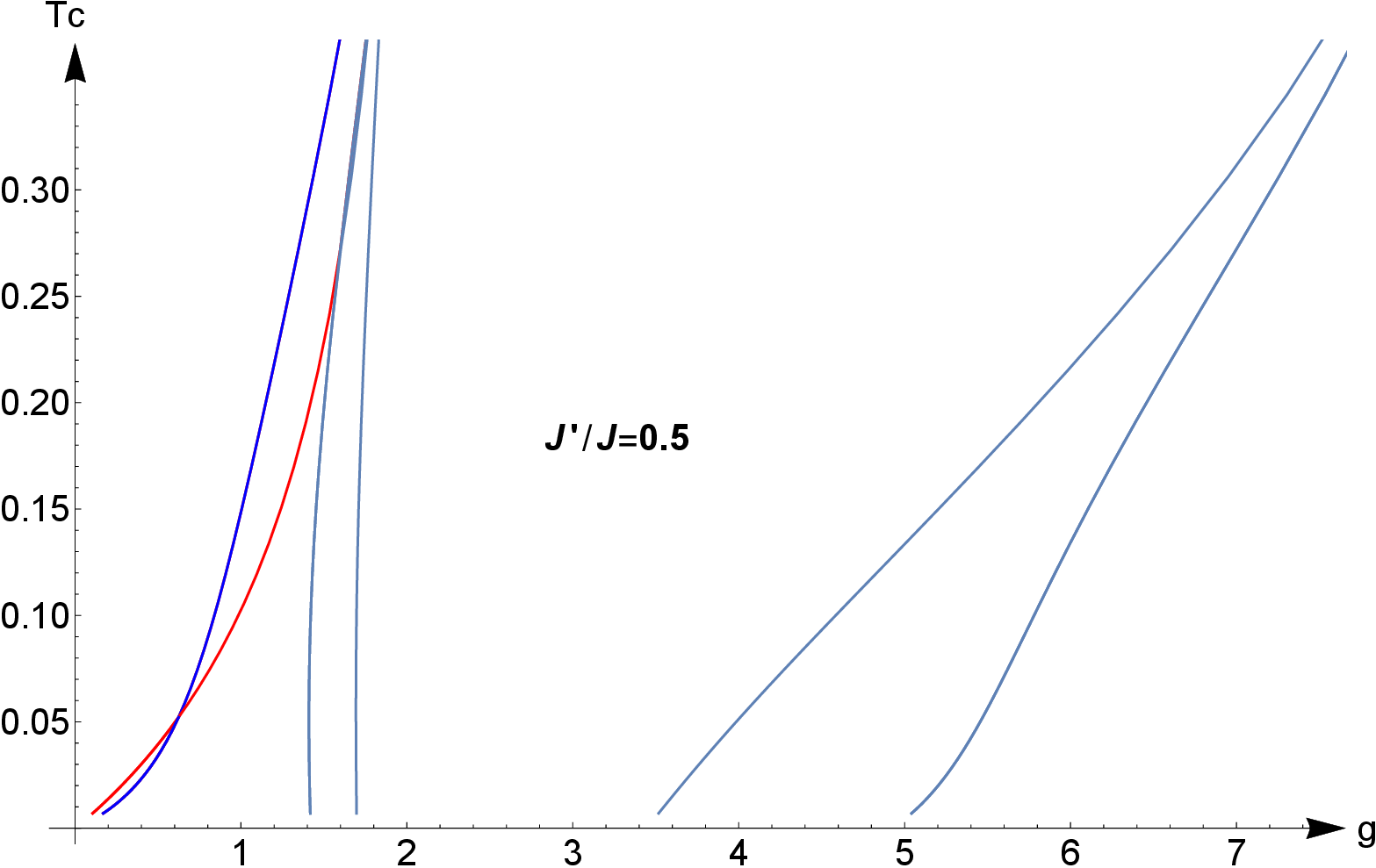}
\includegraphics[width=8cm]{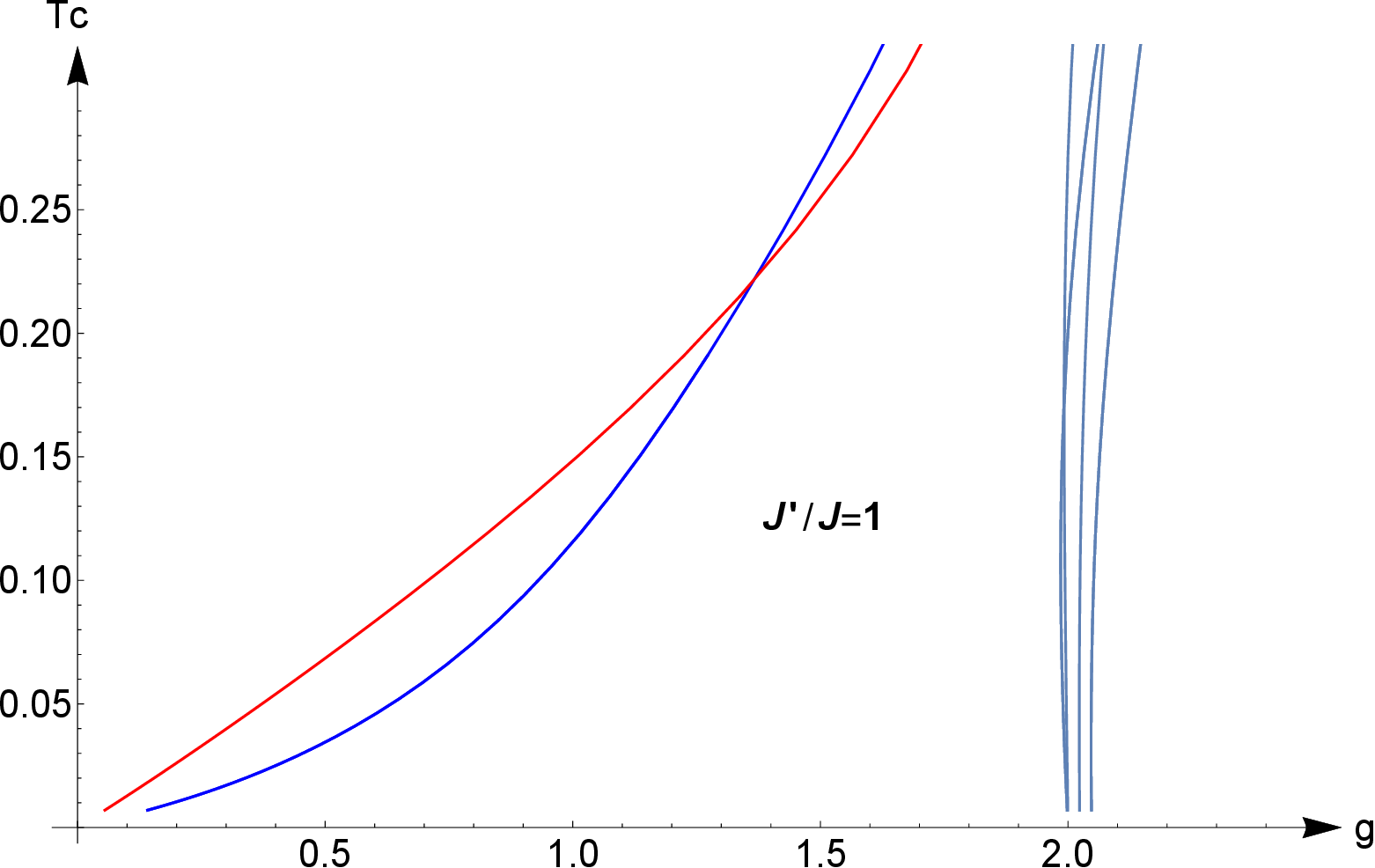}
\includegraphics[width=8cm]{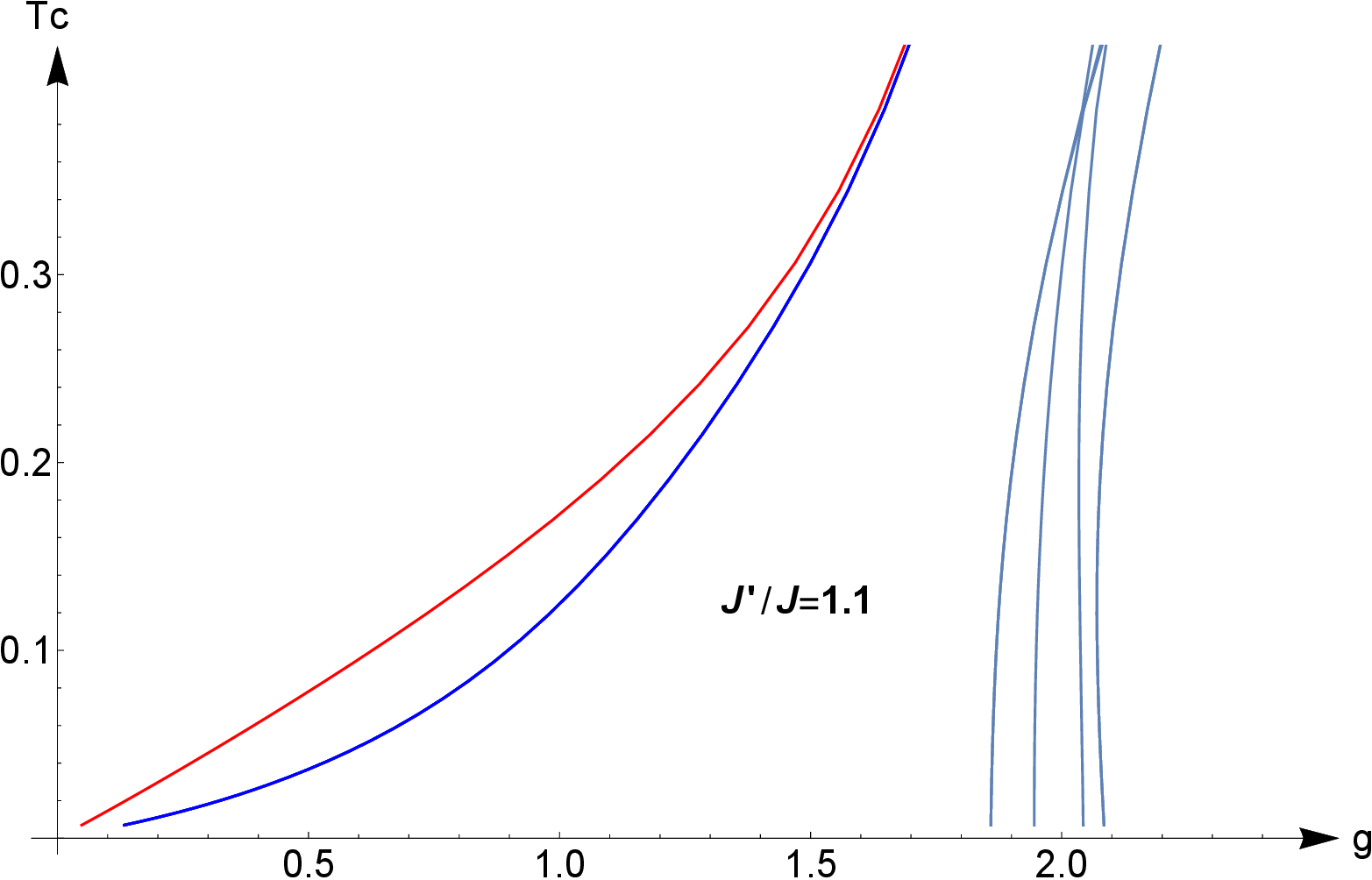}
\includegraphics[width=8cm]{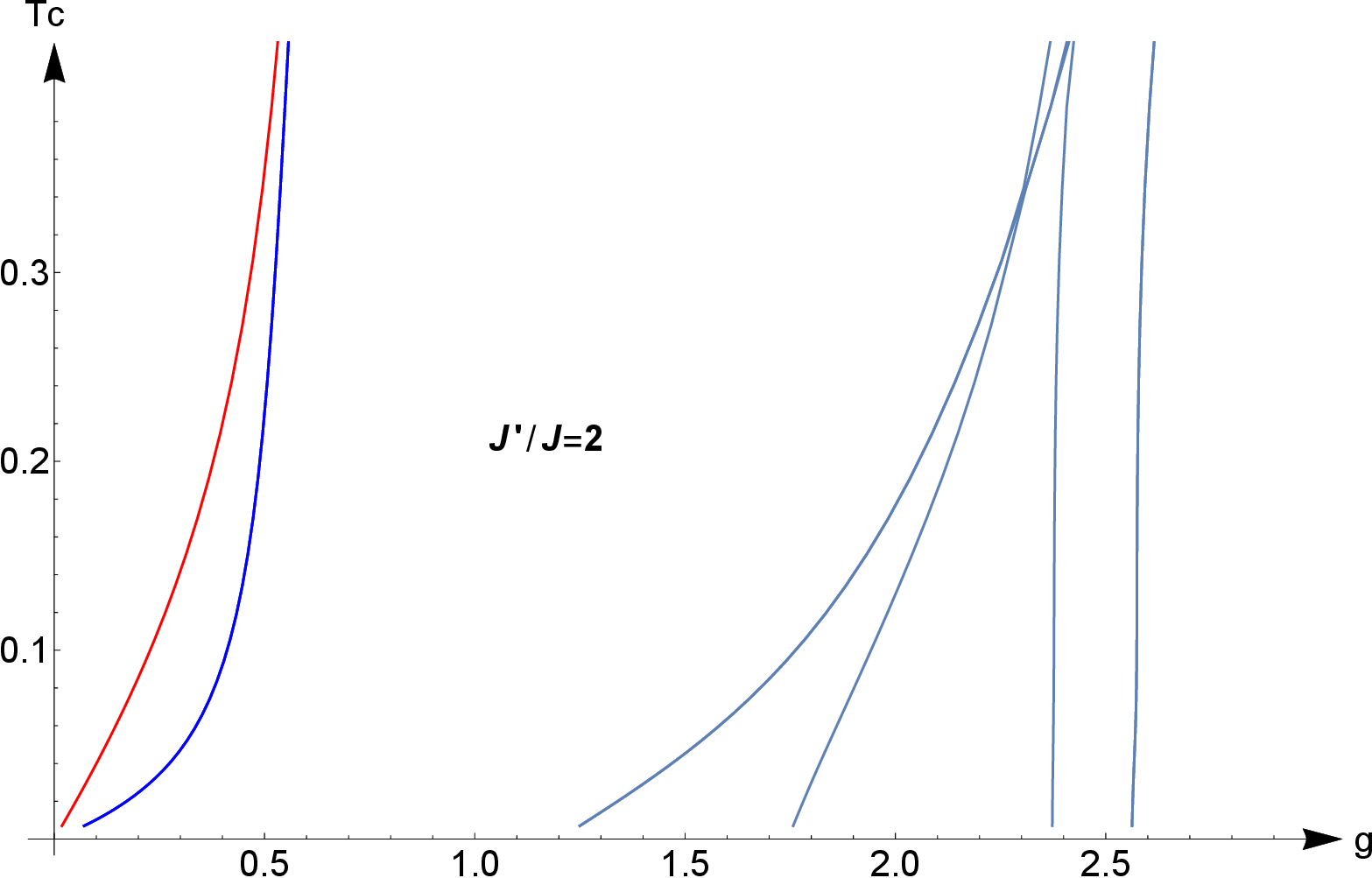}
\caption{Dependence of the critical transition temperatures, $T_c$, with coupling strength in the phenomenological $t-t'-J-J'$ model at hole doping, $\delta=0.15$. 
The critical temperatures associated with the nine superconducting pairing possibilities 
expressed in Eq. (\ref{eq:sc})
are shown at a fixed hole doping of $\delta=0.15$. The $J'/J$ anisotropy 
is varied from $J'/J=0.5$  (top left) to $J'/J=2$ (bottom right). 
The $s^+$ pairing state (red line) denoted by $s^*$,
the doubly degenerate $d^+$ states (dark blue line) denoted by $d^*$, are the most favorable followed by the other
six superconducting states of Eq. (\ref{eq:sc}) which require a stronger coupling (larger $g$) 
to be formed. We have fixed $t \equiv 1$ in all plots.}
 \label{fig:Tcdop015}
 \end{figure*}
 
 \begin{figure*}[!t]
   \centering
\includegraphics[width=7.5cm]{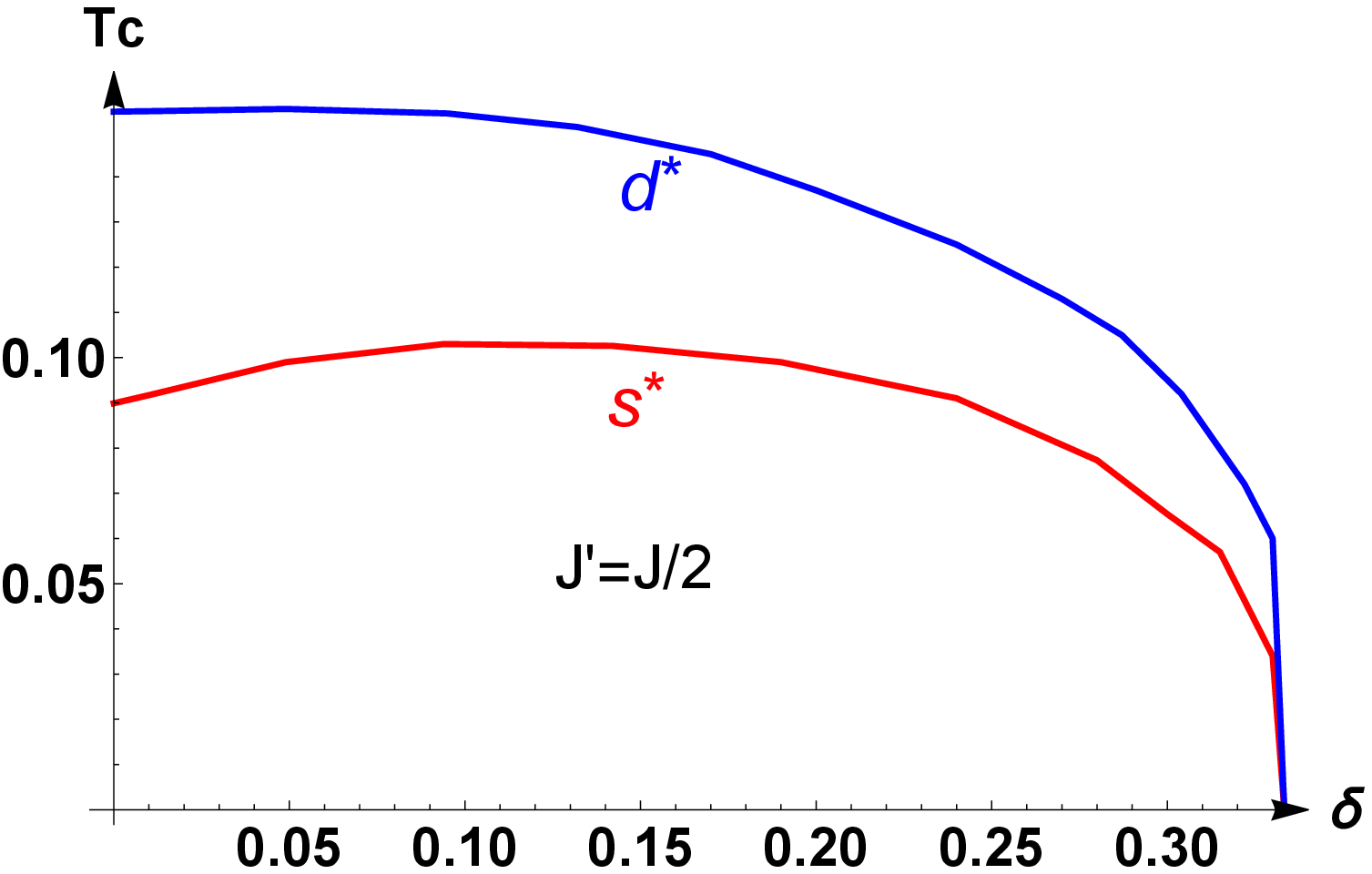}
\includegraphics[width=7.5cm]{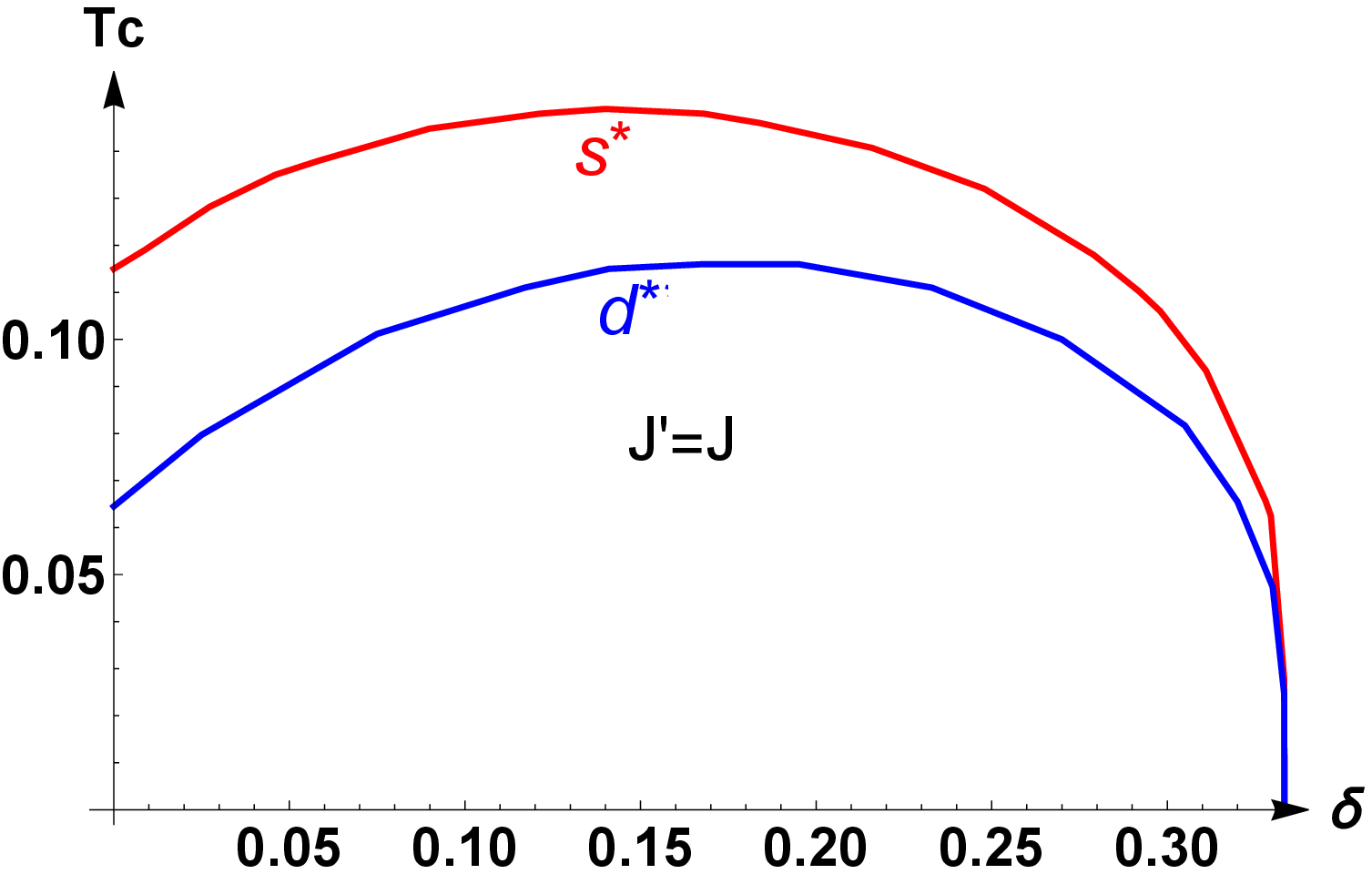}
\caption{Dependence of the critical temperature, $T_c$, with hole doping
in the phenomenological $t-t'-J-J'$ model. While for $J'=0.5J$ the dominant
paring is $d^*$-wave (doubly-degenerate) with the highest $T_c$ of all the 
nine possible pairing states for $J'=J$ it is the $s^*$ state 
that has the highest $T_c$ in the whole doping range analyzed. 
We have fixed $J=t\equiv 1$.}
 \label{fig:phased}
 \end{figure*} 
The matrix elements, $\Omega^{p}_{mn}({\bf k})$ read:
\begin{eqnarray} 
\Omega^1_{mn}({\bf k} )&=&C_{A1}^*(\epsilon_m) C_{A2}(\epsilon_n) +C_{A2}^*(\epsilon_m) C_{A1}(\epsilon_n)
\nonumber \\
\Omega^2_{mn}({\bf k} )&=&C_{A1}^*(\epsilon_m) C_{A3}(\epsilon_n) +C_{A3}^*(\epsilon_m) C_{A1}(\epsilon_n)
\nonumber \\
\Omega^3_{mn}({\bf k} )&=&C_{A2}^*(\epsilon_m) C_{A3}(\epsilon_n) +C_{A3}^*(\epsilon_m) C_{A2}(\epsilon_n)
\nonumber \\
\Omega^4_{mn}({\bf k} )&=&C_{A3}^*(\epsilon_m) C_{B3}(\epsilon_n) e^{i {\bf k} {\delta}_3} +C_{B3}^*(\epsilon_m) C_{A3}(\epsilon_n) e^{-i {\bf k} {\delta}_3}
\nonumber \\
\Omega^5_{mn}({\bf k} )&=&C_{A2}^*(\epsilon_m) C_{B2}(\epsilon_n)e^{i {\bf k} {\delta}_2 }+C_{B2}^*(\epsilon_m) C_{A2}(\epsilon_n)e^{-i {\bf k} {\delta}_2}
\nonumber \\
\Omega^6_{mn}({\bf k} )&=&C_{A1}^*(\epsilon_m) C_{B1}(\epsilon_n) e^{i {\bf k} {\delta}_1}+C_{B1}^*(\epsilon_m) C_{A1}(\epsilon_n)e^{-i {\bf k} {\delta}_1}
\nonumber \\
\Omega^7_{mn}({\bf k} )&=&C_{B1}^*(\epsilon_m) C_{B2}(\epsilon_n) +C_{B2}^*(\epsilon_m) C_{B1}(\epsilon_n)
\nonumber \\
\Omega^8_{mn}({\bf k} )&=&C_{B1}^*(\epsilon_m) C_{B3}(\epsilon_n) +C_{B3}^*(\epsilon_m) C_{B1}(\epsilon_n)
\nonumber \\
\Omega^9_{mn}({\bf k} )&=&C_{B2}^*(\epsilon_m) C_{B3}(\epsilon_n) +C_{B3}^*(\epsilon_m) C_{B2}(\epsilon_n),
\nonumber \\
\end{eqnarray}
where the $C_{\alpha i}(\epsilon_m)$ are the Bloch wave coefficients of the $m$-band, so the symmetries of the pairing amplitudes are determined by these coefficients carrying the information of the symmetries of the lattice.
From the eigenvalues of $\Gamma$, $(1/J)_\lambda$, we obtain the Cooper pair formation energy, $g_\lambda=[(1/J)_\lambda]^{-1}$, 
for each allowed pairing channel, $\lambda=1,...,9$, consistent with the DHL. The smaller $g_\lambda$ means the easier to form the Cooper pairs in the $\lambda$-th pairing channel. 
From the associated eigenvectors {$\Psi_\lambda$} of the $\Gamma$ matrix 
we obtain the possible pairing states. The $\Gamma$ matrix has the
following $3 \times 3$ block structure:
\begin{equation*}
\Gamma=
\begin{pmatrix}
A_{3 \times 3} & B_{3 \times 3}  & C_{3 \times 3}  \\
B_{3 \times 3} & D_{3 \times 3}  & B_{3 \times 3}  \\
C_{3 \times 3} & B_{3 \times 3}  & A_{3 \times 3}  \\
\end{pmatrix}
 \end{equation*}
where:
$$A_{3 \times 3} = \left( \begin{array}{ccc}
\Gamma_{11} & \Gamma_{12} & \Gamma_{12} \\
\Gamma_{12} &  \Gamma_{11} & \Gamma_{12} \\
\Gamma_{12} & \Gamma_{12} & \Gamma_{11} \\
\end{array} \right),
B_{3 \times 3} = \left( \begin{array}{ccc}
\Gamma_{14} & \Gamma_{15} & \Gamma_{15} \\
\Gamma_{15} & \Gamma_{14} & \Gamma_{15} \\
\Gamma_{15} & \Gamma_{15} & \Gamma_{14} \\
\end{array} \right)
$$
$$
C_{3 \times 3} = \left( \begin{array}{ccc}
\Gamma_{17} & \Gamma_{18} & \Gamma_{18} \\
\Gamma_{18} &  \Gamma_{17} & \Gamma_{18} \\
\Gamma_{18} & \Gamma_{18} & \Gamma_{17} \\
\end{array} \right),
D_{3 \times 3} = \left( \begin{array}{ccc}
\Gamma_{44} & \Gamma_{45} & \Gamma_{45} \\
\Gamma_{45} & \Gamma_{44} & \Gamma_{45} \\
\Gamma_{45} & \Gamma_{45} & \Gamma_{44} \\
\end{array} \right).
$$
Since the $A, B, C, D$ matrices above have identical structure, they
share the same three eigenvectors: $\phi^T_{sym}= (1,1,1), (1,-1,0), (1,1,-2) $ which represent the three possible intra-triangular or inter-triangular pairing states: $\phi_s=(1,1,1)$, $\phi_{d_{xy}}=(1,-1,0)$ and $\phi_{d_{x^2-y^2}}=(1,1,-2)$ in the DHL. 
This fact allows to express the 
eigenvalues, $(1/J)_\lambda$, and eigenvectors, {$\Psi_\lambda$}, 
of the $\Gamma$ matrix in terms of the eigenvalues and eigenstates of the $A,B,C,D$ matrices. We find two different types of eigenstates of $\Gamma$ when these are expressed in terms of the eigenstates, $\phi_{sym}$:
\begin{align}
[\Delta_\lambda]^0 & \propto (\phi_{sym}, 0,-\phi_{sym})^T  \nonumber \\
[\Delta_\lambda]^{\pm} & \propto (\phi_{sym}, \beta^{\pm}_{sym} \phi_{sym}, \phi_{sym} )^T
\end{align}
with associated eigenvalues: 
\begin{align}
[(1/J)_\lambda]^0 &= a_{sym}-c_{sym} \nonumber \\
[(1/J)_\lambda]^{\pm} &= a_{sym} +c_{sym}+\beta_{sym}^\pm b_{sym}
\end{align}
where $a_{sym}, b_{sym}, c_{sym}, d_{sym}$ are the eigenvalues of $A$, $B$, $C$, $D$, respectively.
The $\beta_{sym}^\pm$ coefficients are obtained from the eigenvalue problem associated with the $[\Psi_\lambda]^{\pm}$ 
eigenstates which involves the two coupled equations: $[(1/J)_\lambda]^{\pm}=a_{sym} +c_{sym}+\beta_{sym}^\pm b_{sym}$, $[(1/J)_\lambda]^{\pm} \beta_{sym}^\pm=
2 b_{sym}+\beta_{sym}^\pm d_{sym}$. The $\beta_{sym}^\pm$ obtained from these equations read: 
\begin{widetext}
\begin{equation}
\beta_{sym}^\pm={{-(a_{sym} + c_{sym}-d_{sym}) }\pm \sqrt{(a_{sym}+c_{sym}-d_{sym})^2+ 8 b_{sym}^2} \over 2 b_{sym}}.
\end{equation}
\end{widetext}
From the above analysis we finally obtain the nine possible superconducting pairing states in the DHL together with 
their corresponding {Cooper pair formation energies}:
 \begin{widetext}
\begin{eqnarray}
\Delta_f&=&{1 \over \sqrt{6}} [(1,1,1), (0,0,0),(-1,-1,-1)]^T, g={1 \over a_s-c_s}
\nonumber \\
\Delta_{p_x}&=&{1 \over 2 \sqrt{3}} [(1,1,-2), (0,0,0),(-1,-1,2)]^T, g={1 \over a_{d_{x^2-y^2}}-c_{d_{x^2-y^2}}}
\nonumber \\
\Delta_{p_y}&=&{1 \over 2} [(1,-1,0),(0,0,0),(-1,1,0)]^T, g={1 \over a_{d_{xy}}-c_{d_{xy}}}
\nonumber \\
\Delta_{d_{x^2-y^2}^\pm}&=&
{1 \over \sqrt{12 + 6 (\beta_{d_{x^2-y^2}^\pm})^2}}  [(1,1,-2), \beta_{d_{x^2-y^2}}^\pm (1,1,-2), (1,1,-2)]^T, g=J_{d_{x^2-y^2}^\pm}
\nonumber \\
\Delta_{d_{xy}^\pm}&=&{1 \over \sqrt{4 + 2 (\beta_{d_{xy}^\pm})^2}} [(1,-1,0),\beta_{d_{xy}}^\pm (1,-1,0), (1,-1,0)]^T, g= J_{d_{xy}^\pm}
\nonumber \\ 
\Delta_{s^\pm}&=&{1 \over \sqrt{6+3(\beta_s^\pm)^2}} [(1,1,1), \beta_s^\pm (1,1,1), (1,1,1)]^T, g=J_{s^\pm}.
\label{eq:sc}
\end{eqnarray}
\end{widetext}
where we have omitted the $0$ subscript in the island type states {\it i. e.} the $f$, $p_x$ and $p_y$.
The above solutions are sketched out in Fig. 1 of the main text except for the $p$-wave solutions which are not 
encountered in our numerical analysis. 
While the $s^-$ solution with $\beta^-_s>0$ represents a conventional fully gapped $s$-type superconductor, we denote
the $s^+$ as $s^*$ since it has $\beta^+_s<0$ and so it has nodes in the gap. The $d^-$ states which are found to have $\beta^-_d<0$ are conventional $d_{xy}$ and $d_{x^2-y^2}$ states having the two 
corresponding nodal lines in the gap as shown in Fig. 1.
However, the $d^+$ states with $\beta^+_d >0$ are found to have four nodal lines instead, representing an extended version
of the conventional $d$-wave superconducting states and so they are denoted as $d^*$. In our analysis the extended
$s^*$ and $d^*$ states corresponding to the above $s^+$ and $d^+$ are the dominant $s$ and $d$-wave  
superconducting states. Finally, the interesting $f$ wave singlet state also found in the calculations has 
three nodal lines as shown in Fig. 1.

In Fig. \ref{fig:Tcdop015} we show the critical temperatures obtained from the linearized gap equation (\ref{eq:eigen}) at a fixed doping of $\delta=0.15$. Broadly speaking, the $s^+$ pairing state (red line) denoted by $s^*$ (extended $s$) and the doubly degenerate ${d^+}$ states (blue line) cost the least pair formation energy 
since they have the smallest $g$. At this doping, the other six pairing states ($f$, $p_x$, $p_y$, $s^-$ and the doubly-degenerate $d^-$)  cost much more energy occurring at larger $g$'s. We find that while for $J'/J \gtrsim 1.08$, the $s^*$ pairing state is always the most favourable, for $J'/J \lesssim 1.08$ 
the doubly degenerate ${d^+}$-wave solution becomes most favourable.

\begin{figure}
\centering
\includegraphics[width=8.5cm]{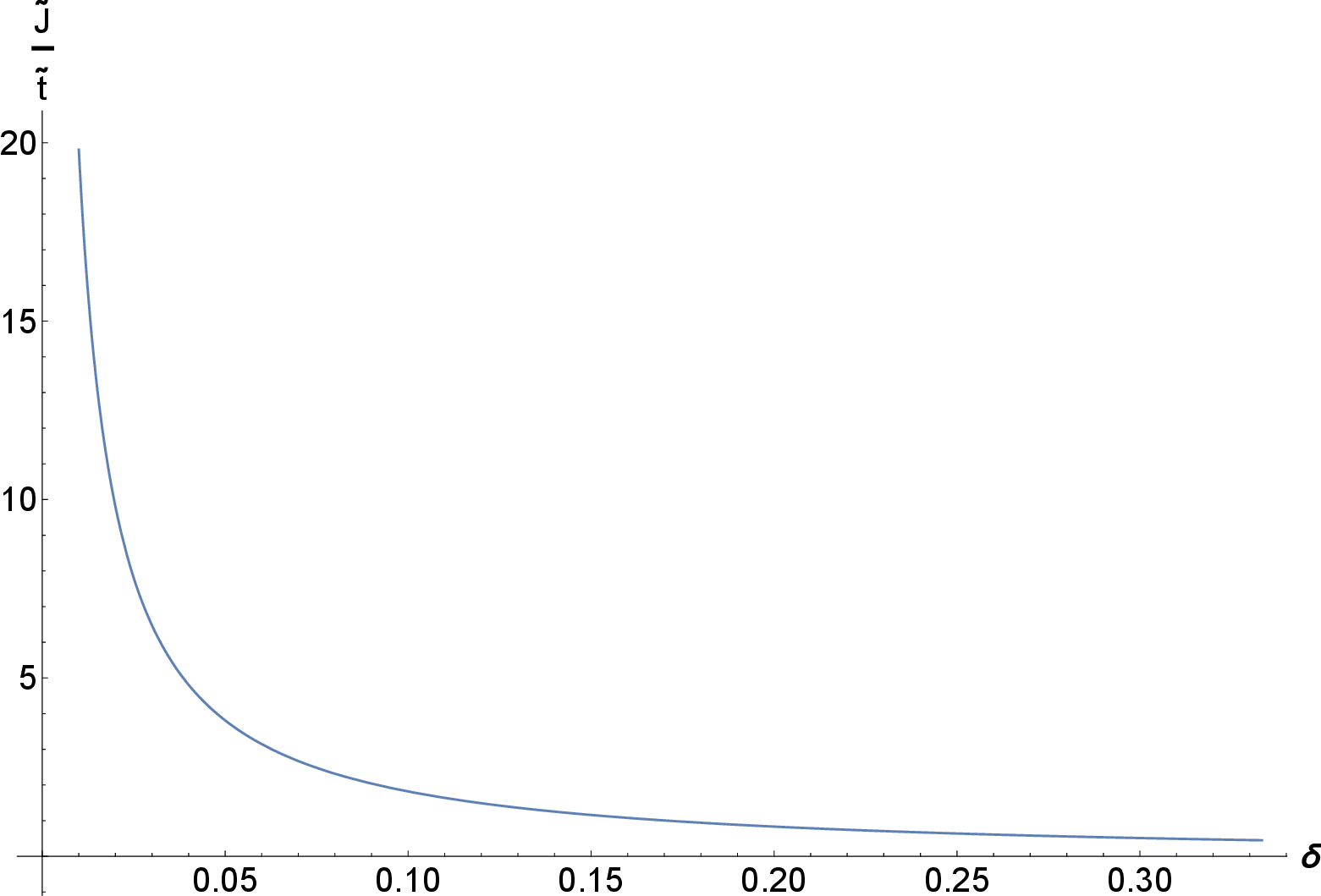}
\caption{Renormalized, $\tilde{J}/\tilde{t}$, ratio as
a function of doping. As $\delta \rightarrow 0$, the ratio
$\tilde{J}/\tilde{t}$ is rapidly enhanced when 
$\delta \rightarrow 0$ due to the double occupancy projection. 
We have taken ${J \over t}=0.1$.}
\label{fig:renorm}
\end{figure}

In order to extend the analysis 
we explore the doping dependence of the most favorable $s^*$ and $d^*$-wave 
states. In Fig. \ref{fig:phased} we show the 
dependence of $T_c$ with $\delta$ for two different $J'/J$ ratios. 
While the $d^*$-wave solution is found to be the most stable one 
with the highest corresponding $T_c$ in the whole doping range 
when $J'/J=0.5$, the $s^*$ is the most stable for $J'/J=1$. This extends
our result for $\delta=0.15$ shown in Fig. \ref{fig:Tcdop015} 
to all doping ranges considered. 

\begin{figure*}
   \centering
\includegraphics[width=7cm]{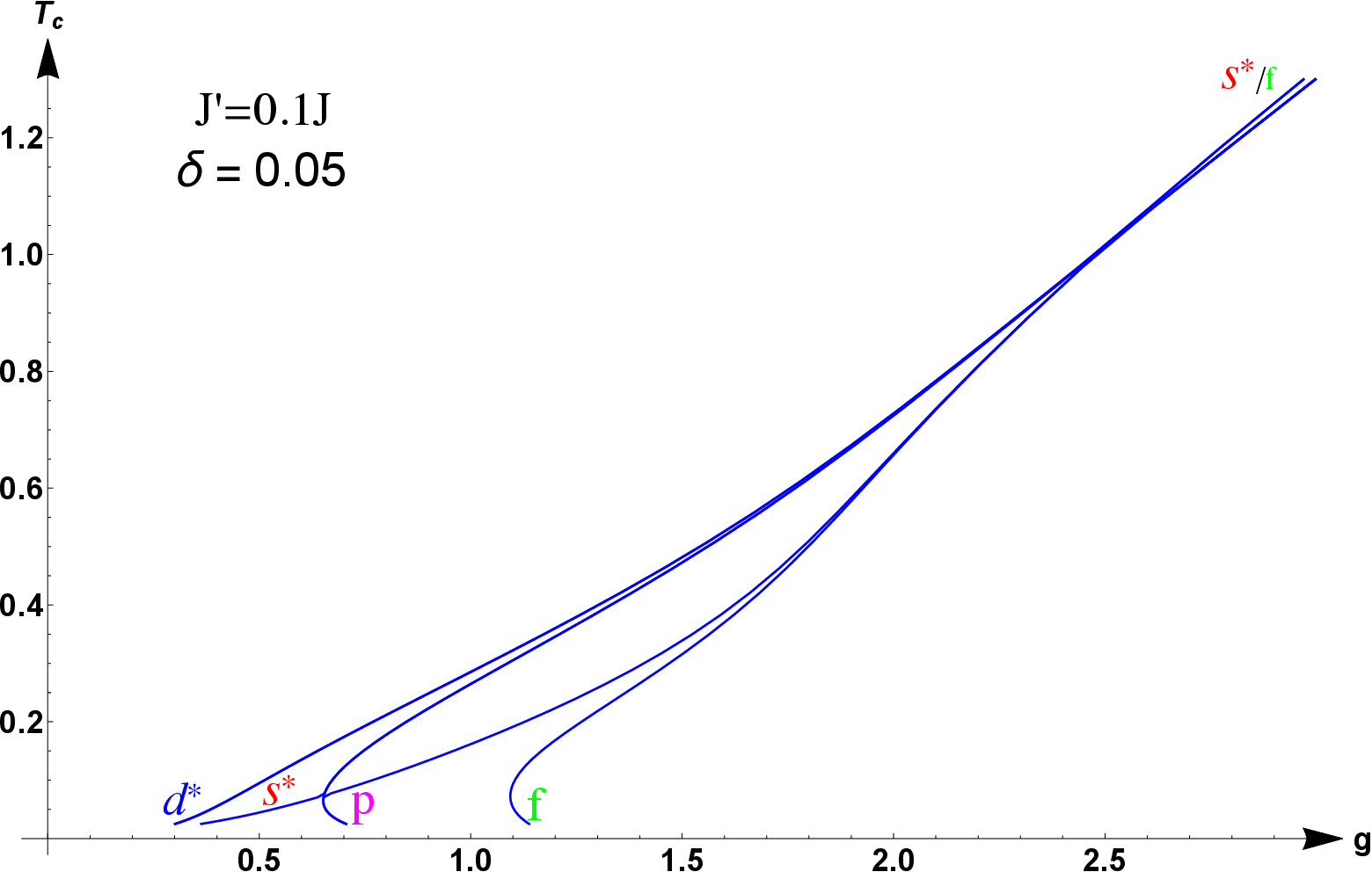}
\includegraphics[width=7cm]{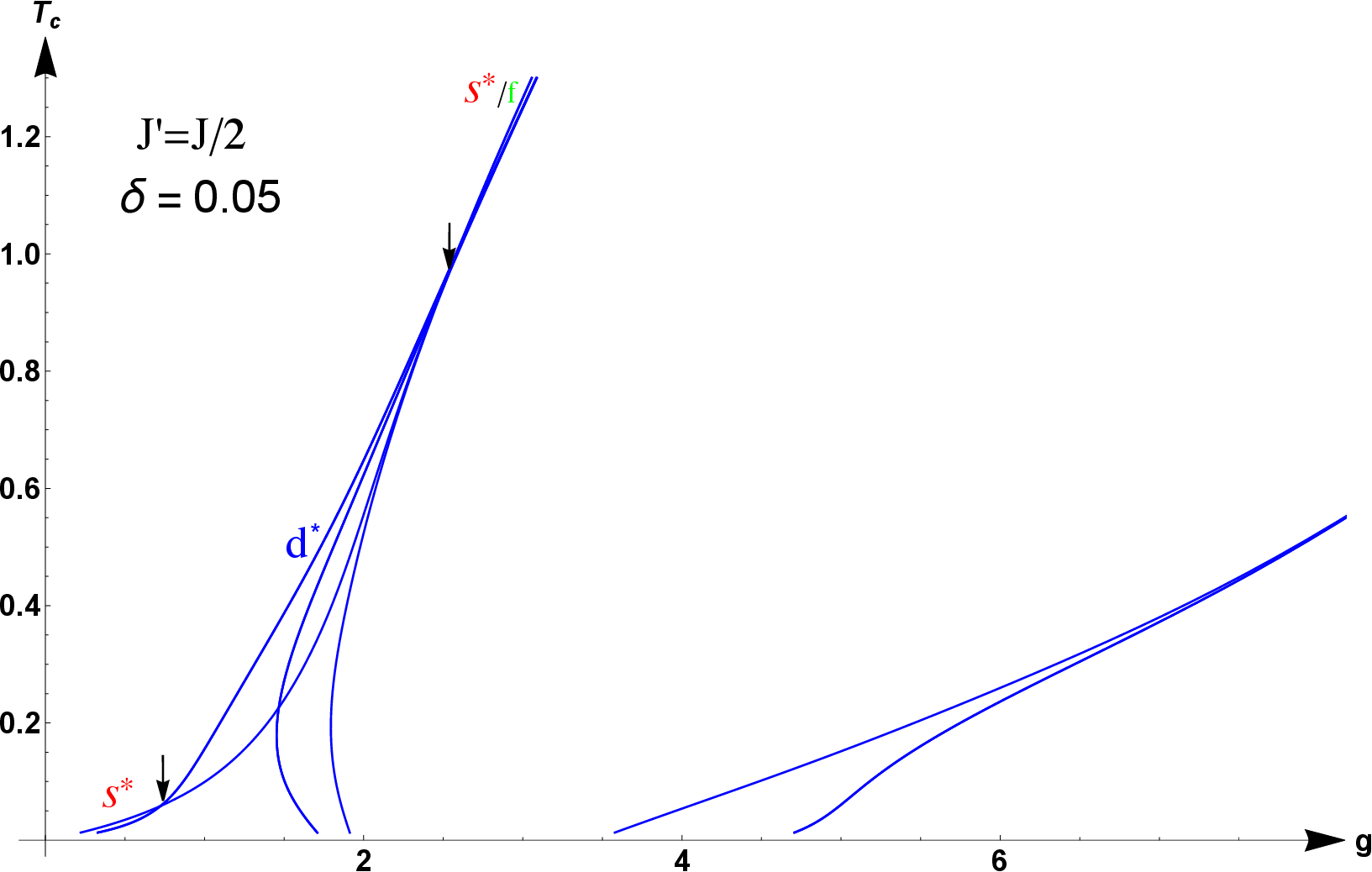}
\includegraphics[width=7cm]{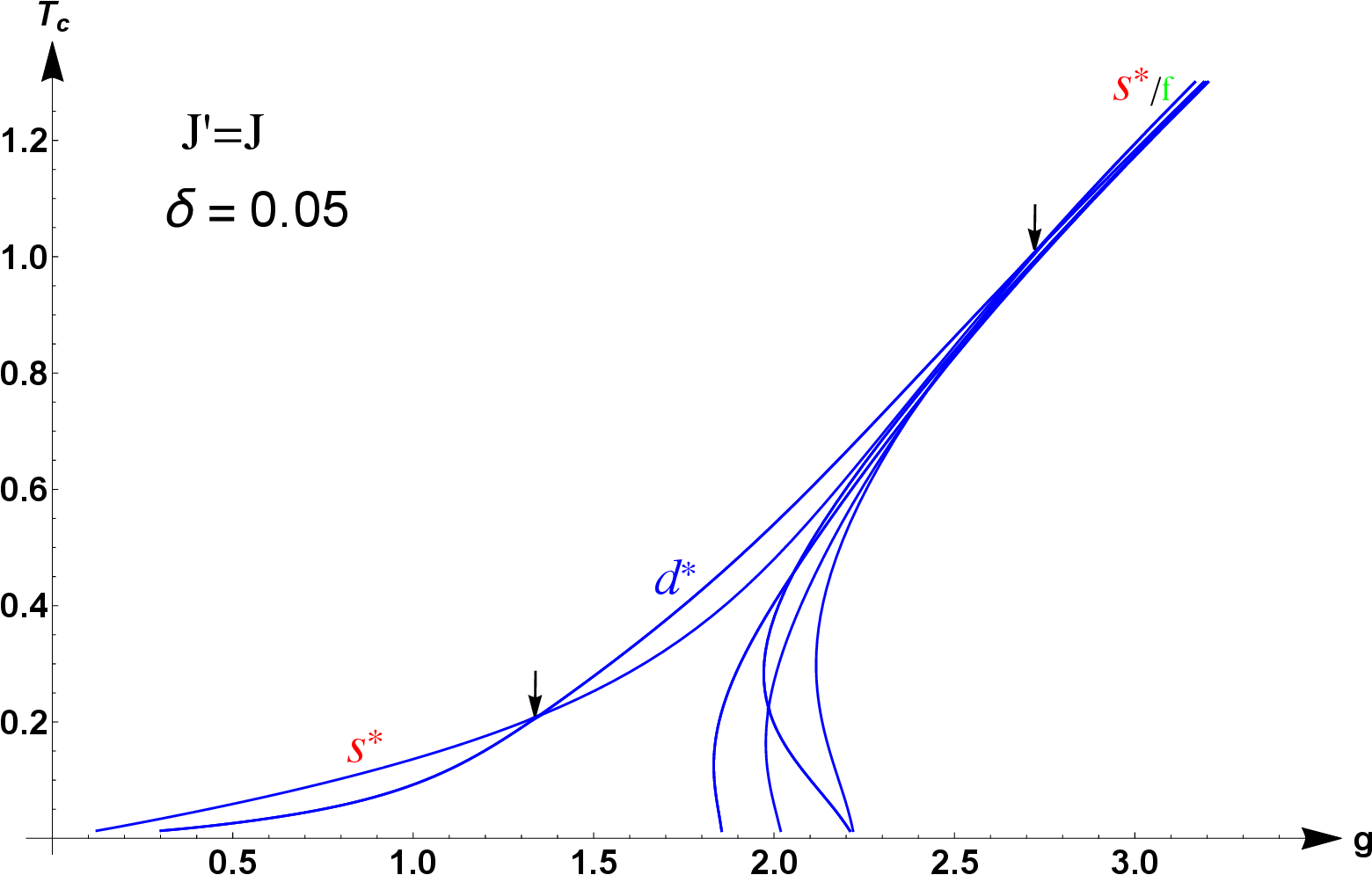}
\includegraphics[width=7cm]{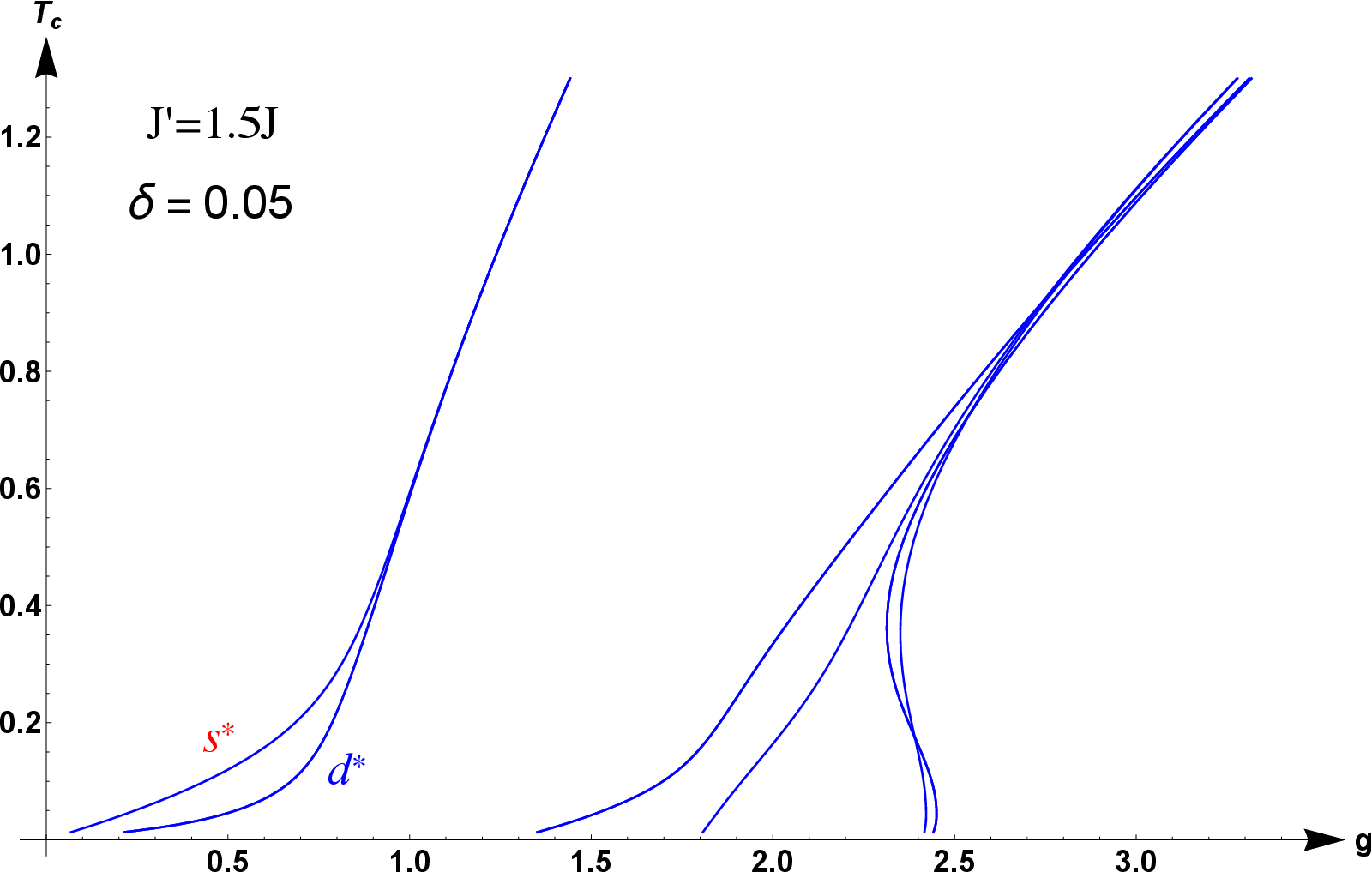}
\caption{Dependence of critical superconducting temperatures 
with coupling strength, $g$, in the phenomenological $t-t'-J-J'$ model with hole doping, 
$\delta=0.05$. The degree of magnetic exchange anisotropy, $J'/J$,
increases from top left to the bottom left panel. Successive switches of 
the pairing symmetry can occur at coupling strengths indicated 
by vertical arrows. For instance, for $J'=J$, a switch from 
$s^*$ to $d^*$-wave occurs first followed by a second switch 
from $d^*$-wave to $s^*$ or $f$-wave symmetry ($s^*/f$ means 
that these two solutions are degenerate) with increasing $g$. 
This can explain the various superconducting solutions observed in 
the phase diagram of the model with double occupancy projected out.  
The case $J'=1.5$ shows how $s^*$ and $d^*$-wave pairing are clearly 
separated from other solutions gradually merging as the coupling strength increases although 
the $s^*$ solution is always favoured.  This 
explains the different superconducting states found when increasing 
$\delta$ as shown in Fig. 1 of the main text.}
\label{fig:Tcdop005}
\end{figure*} 

The phase diagram obtained in Fig. 1 of the main text incorporates 
the projection of the doubly occupied sites i.e., it 
is the phase diagram of model (1) in the main text in which 
double occupancy has been removed from the phenomenological $t-t'-J-J'$
considered above. The main effect of the projection is the renormalization
of the $t, t', J, J'$ parameters of the phenomenological 
hamiltonian. In Fig. \ref{fig:renorm} we show the dependence of the renormalized ${\tilde{J} \over \tilde{t}}=\left( {g_s  \over g_t} \right) 
{J \over t}={2 \over \delta (1+ \delta)} { J \over t}$ with doping for a fixed ${J \over t}=0.1$
(this value is used in the phase diagram of Fig. 1 of the main text). 
The ratios $t'/t, J'/J$ remain fixed to the bare values since
the renormalization of the parameters in the two different bonds is exactly 
the same. As can be observed from Fig. \ref{fig:renorm}, $\tilde{J}/\tilde{t}$ 
increases rapidly with the decrease of $\delta$ with 
 ${\tilde{J} \over \tilde{t}} > 1 $ when $\delta \lesssim 0.17$. 
This means that, in order to corroborate the $f$-wave pairing state found at low $\delta$ 
in the phase diagram of the main text it is important to search for the 
other symmetry allowed solutions to the linearized gap equations of the unprojected
$t-t'-J-J'$ model at low dopings and in the large 
effective coupling, $g= {\tilde{J} \over \tilde{t}}$, regime. 

For this purpose we show in Fig. \ref{fig:Tcdop005} the dependence of the critical temperature, $T_c$ with the pairing coupling strength, $g$, at a fixed 
low doping, $\delta=0.05$, for different $J'/J$. The pairing solutions are 
similar to the ones found for $\delta=0.15$ discussed previously.
At large couplings, above the vertical rightmost arrow, the
$f$-wave solution becomes quasi-degenerate with the $s^*$ solution.
Since as explained above, the coupling of the 
projected model is effectively $g = \tilde{J}/\tilde{t}$,  
which becomes very large at low doping: $g = \tilde{J}/\tilde{t}\approx 3.81$ 
for $\delta=0.05$ with $J/t=0.1$ (see Fig. \ref{fig:renorm} )
$f$-wave pairing is a possible ground state of the $t-t'-J-J'$ model
taking into account the full projection.
Indeed $f$-wave pairing occurs at low hole doping $\delta=0.05$ in the phase diagram 
of Fig. 1 of the main text which is obtained by solving the 
full non-linear set of equations (\ref{eq:sce}). This analysis 
serves as a check of the numerical solution to the equations (\ref{eq:sce})
corroborating the existence of the $f$-wave pairing solution.

\section{Pairing correlations from exact diagonalization on small clusters}

We further explore the unconventional pairing states found in our RVB mean-field theory 
on the $t-t'-J-J'$ model by using exact diagonalization (ED) of the 
appropriate Hubbard model on small clusters. An RVB state has been 
recently found in ED  calculations of the undoped 
Hubbard model \cite{manuel2020} on the DHL. Here, we extend such 
previous ED analysis to elucidate which kind of pairing symmetry 
underlies the RVB state. Hence, we introduce the following pairing correlations \cite{merino2014}:

\begin{widetext}

\begin{equation}
    P_\lambda=\frac{1}{4N_c}\sum_{\langle \alpha i,\beta j\rangle}\sum_{ \langle \gamma k,\delta l\rangle}\Delta_{l}{(\alpha i,\beta j)}\Delta_{l}{(\gamma k,\delta l)}\left(\langle c_{\alpha i\uparrow}^\dagger c_{\beta j \downarrow}^\dagger c_{\gamma k\uparrow}c_{\delta l \downarrow}\rangle -\langle c^\dagger_{\alpha i\uparrow}c_{\gamma k\uparrow}\rangle\langle c^\dagger_{\beta j\downarrow}c_{\delta l\downarrow}\rangle\right)
\label{EDPairings}
\end{equation}
\end{widetext}
where $N_c=6$, $\alpha,\beta,\gamma,\delta =A,B$, $i,j,k,l= 1,2,3$ and $\lambda=1,...,9$. 
The nine-component vectors $\Delta_\lambda$ are described for each pairing symmetry, $\lambda$, in \eqref{eq:sc} but taking the $\beta^{\pm}_{sym}$ coefficients as $\mp 1$ which imposes the corresponding
symmetry.
The correspondence between the vector components of ${\bf \Delta}$ and the 
Cooper paired bonds $(\alpha i, \beta j)$ are indicated in Eq. \eqref{eq:pairs}.
\begin{figure}[!t]
   \centering
\includegraphics[width=8cm]{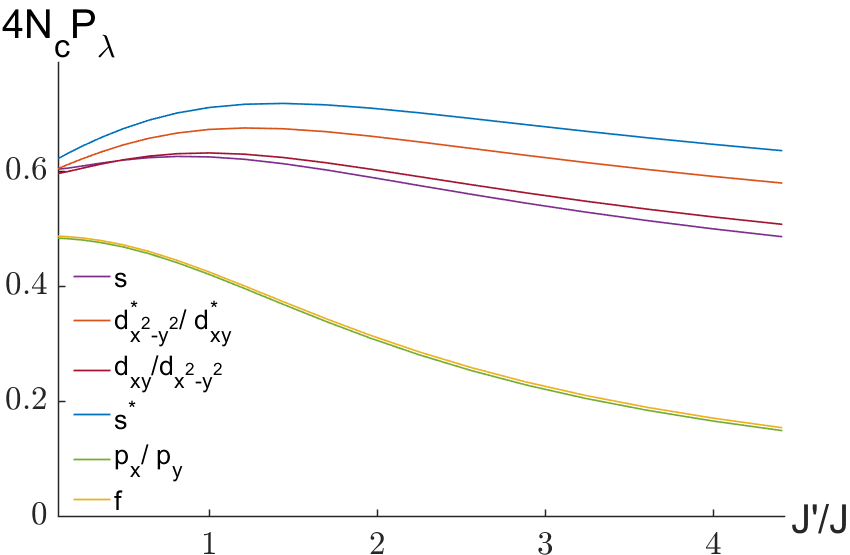}
\caption{Dependence of pairing correlations on $J'/J$. 
Exact diagonalization calculations of $P_\lambda$  
of the half-filled Hubbard model on a six-site cluster 
are shown. The largest pairing correlations in the RVB state occur 
in the extended-$s$ ($s^*$) pairing channel followed by the two degenerate
extended-d, ($d^*$) channels 
in the whole $J'/J$ range . 
We fix $U=20t$. 
}
 \label{fig:pairinged}
 \end{figure} 
Apart from the expected value of the product of the creation and annihilation pairing operators, note that there is a substracted term which suppresses the contribution from the $i=j$ pairing amplitudes. Although these are substantial in small clusters, they are less important for quantifying actual superconducting correlations in the extended lattice.
\begin{figure}[!t]
   \centering
\includegraphics[width=8cm]{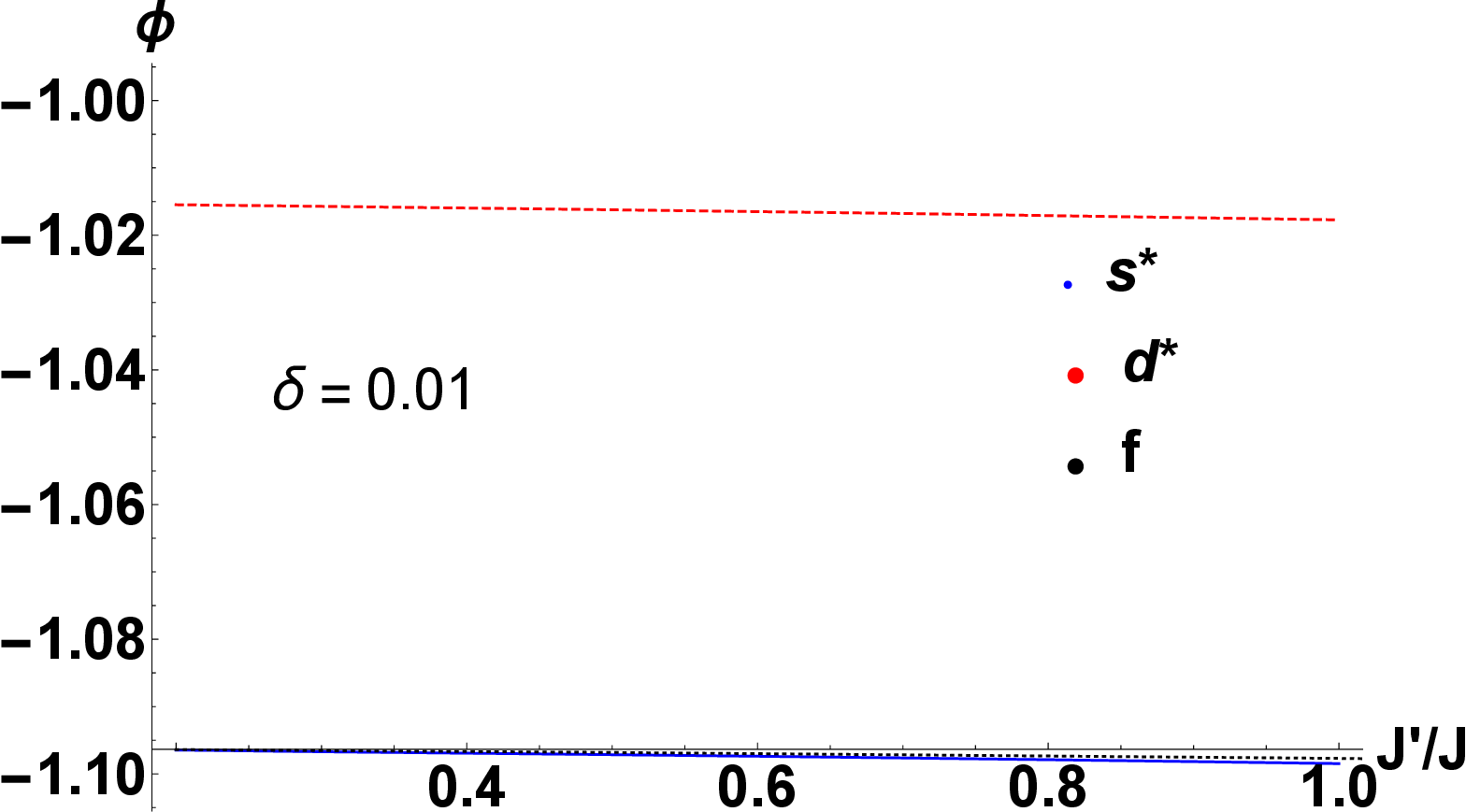}
\caption{Dependence of the free energies of the $s^*$, $d^*$ 
and $f$-wave pairing states on $J'/J$. The free energies are 
obtained from the RVB mean-field treatment of the $t-t'-J-J'$ model 
with $\delta=0.01$ at $T=0.01$. 
}
 \label{fig:freeenergydop0.01T0.01}
 \end{figure} 


The dependence of pairing correlations of a Hubbard model on the DHL with $J'/J$
are displayed in Fig. \ref{fig:pairinged}.  In order to make contact with 
the relevant $t-t'-J-J'$ model we use the second order expression: 
$J(J')=4t^2(t'^2)/U$ with $U=20t$. We find that the extended s-wave ($s^*$) pairing channel dominates in the whole $J'/J$ range while pairing correlations of the $f$, $p_x$, and $p_y$ solutions are always weaker. Furthermore we observe that extended d-wave ($d^*$) pairing dominates over the conventional $d$-wave. These results are consistent with our RVB mean-field treatment of the $t-t'-J-J'$ model. Indeed, as Fig. \ref{fig:freeenergydop0.01T0.01} 
shows the mean-field RVB free energy of the $s^*$ state is the lowest as $\delta \rightarrow 0$ 
in the whole $J'/J$ range.

Finally, we have found that $f$-wave instead of 
$s^*$-wave pairing correlations become the strongest when doping the six-site 
cluster with two holes, $\delta=1/3$. This ED result confirms qualitatively 
that hole doping the DHL can give way to $f$-wave pairing as found with 
RVB mean-field theory. In contrast to these ED results, however, 
$f$-wave pairing arises at a much smaller doping in the RVB mean-field theory, 
around $\delta=0.05$ (see Fig. 1 of the main text). However, our present ED analysis 
does not allow reaching a definitive conclusion on the precise 
doping range of existence of the $f$-wave solution in the 
actual infinite extended lattice due
to the much larger finite size effects expected in small {\em metallic} 
hole doped clusters as compared with {\em insulating} half-filled clusters.

\appendix